\pdfoutput=1
\documentclass[superscriptaddress, 12pt, aps, prl,  preprint, longbibliography, nobibnotes]{revtex4-1}
\usepackage{graphicx}
\usepackage{epstopdf}
\usepackage{color, soul}
\usepackage{amsmath, amssymb}

\begin{document} 

\title{
Unveiling the hidden reaction kinetic network of carbon dioxide in supercritical aqueous solutions
}
\author{Chu Li}
\email{cliag@connect.ust.hk}
\affiliation{Department of Physics, The Hong Kong University of Science and Technology, Hong Kong, China}
\author{Yuan Yao}
\affiliation{Department of Mathematics, The Hong Kong University of Science and Technology, Hong Kong, China}
\author{Ding Pan}
\email{dingpan@ust.hk}
\affiliation{Department of Physics, The Hong Kong University of Science and Technology, Hong Kong, China}
\affiliation{ Department of Chemistry, The Hong Kong University of Science and Technology, Hong Kong, China}
\affiliation{HKUST Shenzhen-Hong Kong Collaborative Innovation Research Institute, Shenzhen, China}

\date{\today}

\begin{abstract}
Dissolution of CO$_2$ in water followed by the subsequent hydrolysis reactions is of great importance to the global carbon cycle, and carbon capture and storage. 
Despite enormous previous studies, 
the reactions are still not fully understood at the atomistic scale. Here, we combined \textit{ab initio} molecular dynamics simulations with Markov state models to elucidate the reaction mechanisms and kinetics of CO$_2$ in supercritical water both in the bulk and nanoconfined states.
The integration of unsupervised learning with first-principles data allows us to identify complex reaction coordinates and pathways automatically instead of \textit{a priori} human speculation.
Interestingly, our unbiased modelling found a novel pathway of dissolving CO$_2$(aq) under graphene nanoconfinement, involving the pyrocarbonate anion (C$_2$O$_5^{2-}$(aq)) as an intermediate state.
The pyrocarbonate anion was previously hypothesized to have a fleeting existence in water; however our study reveals that it is a crucial reaction intermediate and stable carbon species in the nanoconfined solutions. We even observed the formation of pyrocarbonic acid (H$_2$C$_2$O$_5$(aq)), which was unknown in water. 
The unexpected appearance of pyrocarbonates is related to the superionic behavior of the confined solutions. 
We also found that carbonation reactions involve collective proton transfer along transient water wires, which exhibits concerted behavior in the bulk solution but proceeds stepwise under nanoconfinement.
Our study highlights the importance of large oxocarbons in aqueous carbon reactions, with great implications for the deep carbon cycle and the sequestration of CO$_2$.

\end{abstract}

\maketitle
\section{Significance Statement}
Carbon-bearing aqueous solutions play important roles in carbon transport and storage and CO$_2$ mineralization in Earth's interior. However, current understanding of the dissociation of CO$_2$ and subsequent hydrolysis reactions at the molecular scale remains relatively limited, especially when considering elevated pressure and temperature conditions, as well as nanoconfined environments. Here we developed an unbiased first-principles simulation method to elucidate the reaction kinetics and discovered the pyrocarbonate anion (C$_2$O$_5^{2-}$(aq)) as a pivotal reaction intermediate and a stable carbon species in the nanoconfined solutions. This finding challenges previous studies which suggested the instability of C$_2$O$_5^{2-}$(aq)  in aqueous environments. Our study emphasizes the significance of large oxocarbons in water, which has profound implications for the deep carbon cycle and the CO$_2$ sequestration.

\section{Introduction}
Aqueous carbon reactions are integral to Earth's carbon cycle, with significant implications for the carbon budget, climate, and the overall biosphere \cite{falkowski2000global, manning2013chemistry, sverjensky2014important}. 
Notably, aqueous carbon solutions not only exist in the bulk phase but can also be absorbed within nanoscale cavities or nano-spaces \cite{gautam2017structure, marquardt2018structure, huang2020experimental, stolte2022nanoconfinement}.
In deep Earth, the mineral materials comprise nanoscale pores and grain boundaries, significantly influencing the chemical and transport properties of aqueous solutions. The dissolved carbon in water plays a critical role in the carbon transport within the deep carbon cycle, which connects the carbon reservoirs in Earth's surface and interior. 
In CO$_2$ capture and sequestration, carbonation reactions leading to CO$_2$ mineralization typically occur within nano- or sub-nanoscale water films confined by rocks, holding promise for effectively stabilizing and storing CO$_2$ in Earth's subsurface \cite{matter2009permanent, matter2016rapid, sanna2014review, snaebjornsdottir2020carbon}. 
Understanding the carbon reactions in water in a large pressure-temperature (P-T) range 
has garnered great interest among scientists from diverse fields, including chemistry, engineering, Earth and environmental sciences \cite{stirling2010h2co3, manning2013chemistry, facq2014situ, lam2015hydration,  pan2016fate, Abramson2017a,dasgupta2023hydrophobic}.  
Nevertheless, our current understanding of the reaction mechanisms and pathways at the molecular level remains rather limited.

Previous geochemical models assumed that molecular CO$_2$(aq) is the major carbon species in aqueous geofluids, but recent theoretical and experimental studies showed that most of CO$_2$(aq) reacts with water to become CO$_3^{2-}$ and HCO$_3^-$ anions and even H$_2$CO$_3$(aq) in Earth's upper mantle \cite{facq2014situ, pan2016fate, Abramson2017a}. 
As for larger oxocarbons, the pyrocarbonate ion was hypothesized to explain the NMR experiment in aqueous carbonate solutions \cite{zeller2005hidden}, but was only detected in molten carbonates in the absence of water \cite{zhang2013first,manning2013chemistry,corradini2016carbon}. 
It is very challenging to experimentally study the dissolution of CO$_2$(aq) and its subsequent hydrolysis reactions at high P and high T conditions.
Classical and \textit{ab initio} molecular dynamics (AIMD)
simulations
have been widely used 
\cite{stirling2010h2co3, stirling2011hco3, stolte2019large,stolte2022nanoconfinement,dasgupta2023hydrophobic}.
AIMD is particularly advantageous for simulating chemical reactions under extreme conditions,
because the breaking and forming of chemical bonds can be described at the quantum mechanics level.
AIMD does not require the assignment of force field parameters to carbon species, eliminating the need for prior knowledge of possible reaction intermediates and products, which are often difficult to know under extreme conditions.
To obtain reaction pathways and kinetics, many previous studies employed enhanced sampling methods, which rely on predefined reaction coordinates or collective variables (e.g., \cite{stirling2010h2co3, stirling2011hco3, dasgupta2023hydrophobic}).
It is worth noting that the reaction coordinates may oversimplify the complexity of complete reaction networks and the biased sampling may cause the loss of the kinetic information.

The Markov state model (MSM), which represents continuous dynamics as a sequence of Markovian transitions among partitioned spaces at discrete time intervals, enables the elucidation of complex reaction kinetics without any predefined reaction coordinate \cite{prinz2011markov, husic2018markov, wang2018constructing, konovalov2021markov}.
This method automatically identifies the reaction coordinates and parallel kinetic pathways with multiple intermediates using unsupervised machine learning approaches rather than human speculation on reaction mechanisms. 
Besides, the MSM constructed from unbiased MD simulation data maintains the fidelity of kinetics, 
while other chemical reaction network methods usually applied empirical relations to obtain reaction kinetics \cite{unsleber2020exploration, wen2023chemical}.
The MSM has been mainly used to predict long-timescale dynamics of biomolecules by combining short dynamics, such as protein folding and ligand-receptor binding \cite{bowman2011atomistic, da2016bridge, plattner2015protein}. It has also been applied to investigate molecular self-assembly kinetics \cite{tang2017construction, zheng2017kinetics, weng2020revealing} and dominant pathways in ice nucleation \cite{li2021temperature}. However, 
those previous studies utilizing MSMs are based on classical MD simulations, 
in which the predesigned force fields may limit the reaction configuration space.

Here, we applied unsupervised machine learning techniques to construct the MSMs from parallel AIMD simulations to study the reaction kinetic networks of CO$_2$ in supercritical water both in bulk and under graphene nanoconfinement. 
To the best of our knowledge, it is the first time to combine Markov state modelling with AIMD simulations to analyze the kinetics of aqueous chemical reactions.
In the bulk solution, we observed the formation of pyrocarbonate ions that rapidly dissociate. 
However,
our method reveals their unexpected stability under nanoconfinement.
Interestingly, in contrast to the direct carbonation reaction of CO$_2$(aq) with H$_2$O or OH$^-$ in the bulk solution, nanoconfinement facilitates a new reaction pathway involving pyrocarbonate ions as intermediates. We even observed the formation of the hypothetical pyrocarbonic acid.
The unexpected stability of the pyrocarbonic acid and its derivatives is related to the superionic behavior of the confined solutions.
We also found that the carbonation reactions involve collective proton transfer along transient water wires, which exhibits concerted behavior in the bulk solution but proceeds stepwise under nanoconfinement.
Modulating the 2D confinement can be a highly effective strategy for regulating aqueous chemical reactions.
Our findings suggest that large oxocarbons, which were previously ignored, may be an important carbon host in the deep carbon cycle, and play an important role in the sequestration of CO$_2$.
The combination of quantum molecular dynamics, Markov state models, and unsupervised machine learning techniques demonstrates immense potential in unraveling intricate reaction pathways and kinetics.

\section{Results and Discussion}
\subsection {Constructing Markov state models}
The MSM approach treats continuous dynamics as a series of Markovian transitions within partitioned space, where the probability of transitioning to the next state depends solely on the current state.
Here, we aim to simulate the reaction kinetics of carbon in supercritical water as Markovian transitions among discrete partitions of its reaction space. 
Before constructing MSMs, we performed a series of parallel AIMD simulations for CO$_2$ dissolved in the bulk and nanoconfined water at $\sim$10 GPa and 1000 K;
this P-T condition is typically found at the bottom of Earth's upper mantle.
In recent years, the synthesis and manipulation of 2-dimensional nanostructures has advanced significantly, making it feasible to control and investigate the behaviors of nanoconfined aqueous solutions precisely. 
The confinement widths of slit pores can now be reduced to less than 1 nm, drawing considerable attention within the related scientific community \cite{munoz2021confinement}.
We confined the solutions using a distance-dependent potential, which was fitted to the interaction energies between graphene and water or CO$_2$ obtained from the simulations of diffusion quantum Monte Carlo \cite{brandenburg2019physisorption} and van der Waals density functional theory \cite{takeuchi2017adsorption}.
There are no chemical reactions between interfaces and solutions, and we mimic hydrophobic confinement.
This approach has been used in many previous theoretical studies, e.g., \cite{chen2016two, munoz2017nanoconfinement, kapil2022first, stolte2022nanoconfinement, hou2024raman}. 
We compared the bulk, bilayer and monolayer solutions as shown in Fig. \ref{system} 
(See Supplementary Methods and Supplementary Table I for more simulation details).

To construct MSMs effectively, it is important to select proper input features that can capture the possible metastable states, which represent the reaction space to explore.
In this study, our primary focus is carbon speciation, 
so we chose to use the states of the three nearest neighboring oxygen (O$_j$) atoms of each carbon atom (C$_i$) to characterize carbon species, inspired by the message passing in neural networks \cite{gilmer2017neural}.
Fig. \ref{framework} shows the schematic diagram outlining the process of constructing the MSM for carbon in the bulk solution.
We input the local coordinates ($\{D_{C_iO_j}\}$) of the surrounding atoms of O$_j$, which is one of the three oxygen atoms closest to C$_i$, as shown in Fig. \ref{framework}a,
and then we applied principal component analysis (PCA) to reduce the dimensionality of $\{D_{C_iO_j}\}$. Based on the explained variance associated with each principal component (PC), as shown in Supplementary Fig. 1,  we kept the top two PCs, which are distinct from the remaining components. The projection of PCA on these two PCs is shown in Fig. \ref{framework}b, where PC1 and PC2 characterize the bonding of O$_j$ with hydrogen atoms and C$_i$, respectively. Particularly, with increasing PC1,  O$_j$ changes from unbonded to bonded states with its closest hydrogen atom, and with increasing PC2, the C$_i$=O$_j$ bond breaks. 
After PCA, we applied the \textit{k}-means clustering method \cite{likas2003global} to partition the PCA distribution into a number of clustered centers, as shown by the blue dots in Fig. \ref{framework}b. Subsequently, in Fig. \ref{framework}c we assigned a microstate to represent C$_i$ as the union set of three cluster centers of its neighboring O$_j$. 
Due to the C3 rotation symmetry of the \textit{sp}$^2$ carbon studied here, thermal fluctuations change the distance order of the three C=O bonds frequently, so we did not sort the C=O bonds by distance.
Our method allows delivering the atomic coordinate information surrounding O$_j$ to C$_i$ effectively and avoids the potential factitious kinetics due to the fluctuations of C=O bond distances. 
Our feature engineering approach is different from those commonly used in biomolecule applications, because in our AIMD simulations, molecules and ions constantly exchange atoms, giving rise to the permutation problem.
We further grouped the microstates into several macrostates using the Robust Perron Cluster Clustering Analysis (PCCA+) \cite{deuflhard2005robust}, so chemical reactions can be understood using existing chemistry knowledge (see Methods and Supplementary Methods for more detailed procedure).

\subsection {Reaction kinetic network of carbon in the bulk solution}
Fig. \ref{MSM-bulk} shows the reaction kinetics derived from our MSM for carbon in the bulk solution.  
Each state in Fig. \ref{MSM-bulk}a represents a macrostate in the MSM, which is a metastable or stable state located in a free energy basin. The macrostates are separated by free energy barriers and connected by potential transition states. 
Our MSM provides a comprehensive description of aqueous carbon reactions. 
In Supplementary Fig. 3, we compared the carbon speciations obtained from our MSM and a long AIMD trajectory \cite{stolte2022nanoconfinement}, and the results are in excellent agreement.
Specifically, the transitions between states i and ii, and ii and iii in Fig. \ref{MSM-bulk}a describe the diffusion of CO$_2$(aq) as it interacts with other molecules. 
The transition between states iv and v is the interconversion of HCO$_3^-$ and H$_2$CO$_3$(aq),
while the transition between the states v and vi is the interconversion of H$_2$CO$_3$(aq) and H$_3$CO$_3^+$.
These carbon species differ in the number of bonded protons.

Importantly, our MSM is able to capture the rate-limited transition between states iii and iv, which represents a reversible reaction:
\begin{equation}
 \mathrm{CO}_2(aq)+\mathrm{H}_2\mathrm{O} \rightleftharpoons \mathrm{HCO}_3^-+\mathrm{H}^+.
\label{transition1}
\end{equation}
The forward reaction is mainly mediated by the proton transfer along a hydrogen-bonded water wire 
as shown in Fig. \ref{MSM-bulk}b.  The backward reaction occurs stochastically through the breaking of the C-O bond.
The mean first passage time (MFPT) from our MSM provides the forward and backward transition times as 126.02 ps and 105.98 ps, respectively, indicating that the forward and backward reactions are equally likely to happen. 
They are much slower than the protonation or deprotonation reaction between states iv and v, or v and vi. (see Supplementary Table II for all the MFPTs).
Our MSM suggests that HCO$_3^-$ is an intermediate in the formation of H$_2$CO$_3$(aq), which aligns with the previous study using metadynamics, despite at the differing ambient conditions \cite{stirling2010h2co3}. 

Remarkably, our MSM also uncovered a hidden reaction intermediate: the pyrocarbonate ion (C$_2$O$_5^{2-}$(aq)). 
For example, the hydrogen pyrocarbonate ion was generated through the formation of the C-O bond between CO$_2$(aq) and HCO$_3^-$:
\begin{equation}
\mathrm{CO}_2 (aq)+\mathrm{HCO}_3^- \rightleftharpoons \mathrm{HC}_2\mathrm{O}_5^-(aq),
\label{transition2}
\end{equation}
Fig. \ref{MSM-bulk}c shows a snapshot for reaction (\ref{transition2}).
We grouped pyrocarbonate ions with different numbers of bonded protons, such as hydrogen pyrocarbonate ion (HC$_2$O$_5^{-}$(aq)) and pyrocarbonic acid (H$_2$C$_2$O$_5$(aq)), into a single macrostate due to the significantly faster rates of protonation and deprotonation than the formation of C$_2$O$_5^{2-}$(aq), and refer to them as pyrocarbonate ions for convenience's sake. The formation of pyrocarbonate ions is considered a rare event due to the relatively long MFPTs compared to the diffusion of CO$_2$(aq), as shown in Fig. \ref{MSM-bulk}d.
Furthermore, the MFPTs for the formation and dissociation of pyrocarbonate ions are very different, indicating the thermodynamic instability of C$_2$O$_5^{2-}$ in aqueous solutions. This observation is consistent with previous studies, where pyrocarbonate ions were detected
in molten carbonates in the absence of water \cite{manning2013chemistry, corradini2016carbon}. 

By analyzing the stationary population of each macrostate in Fig. \ref{MSM-bulk}a, we can determine the equilibrium fractions of carbon species.
These fractions are consistent with mole percents obtained by the long AIMD simulations in the previous study \cite{stolte2019large}, indicating the accuracy of our MSM in predicting long-time dynamics with chemical reactions.
Furthermore, the MSM allows us to simulate reactions and monitor the evolution of carbon species with arbitrary molecular concentrations. Fig. \ref{MSM-bulk}e shows a typical evolution of various carbon species by initializing all the carbon as CO$_2$(aq) (corresponding to state iii in Fig. \ref{MSM-bulk}a). As time evolves, the reactions occur as the carbon atoms transition between different states within the reaction network. Eventually, all the carbon atoms reach a dynamic equilibrium, and the fractions of carbon species stabilize (see Supplementary Methods for more detailed information).

\subsection {Nanoconfinement enhances the formation of pyrocarbonate ions }
Using the same method, we constructed the MSMs to investigate the reaction kinetics of carbon in the bilayer and monolayer thick solutions under confinement, 
as shown in Fig. \ref{MSM-confinement}a and b, respectively. 
Our MSMs reveal that, in general, the reactions under nanoconfinement are enhanced, as evidenced by the reduced MFPTs for most state transitions (see Supplementary Tables II, III, and IV). This can also be illustrated by the faster evolution of carbon species in the nanoconfined solutions, as shown in Supplementary Figs. 14 and 15 compared to Fig. \ref{MSM-bulk} e.  For instance, the MFPT for the carbonation process in reaction (\ref{transition1}) (iii $\longrightarrow$ iv) decreases from 126.02 ps in the bulk solution to 77.66 ps and 36.64 ps under the bilayer and monolayer confinements, respectively.
The observed enhancement of aqueous carbon reactions is in line with previous MD simulations conducted under similar \cite{stolte2022nanoconfinement} or ambient P-T conditions \cite{dasgupta2023hydrophobic}.
Although our MSMs were constructed at the elevated P-T condition, the revealed
reaction mechanisms can still provide valuable insights into the dissolution of CO$_2$(aq) and the subsequent hydrolysis reactions at near ambient conditions.

Surprisingly, our MSMs show that the unstable pyrocarbonate ions in the bulk solution become stable under nanoconfinement. Fig. \ref{MSM-confinement}c shows a significant increase in the mole percent of pyrocarbonate ions, rising from 0.39\% in the bulk solution to 8.2\% and 27.6\% under the bilayer and monolayer confinements, respectively. 
The pyrocarbonic acid and its derivatives frequently convert into each other through protonation and deprotonation reactions:
\begin{equation}
\mathrm{H}_2\mathrm{C}_2\mathrm{O}_5(aq) \rightleftharpoons \mathrm{HC}_2\mathrm{O}_5^-(aq) + \mathrm{H}^+ \rightleftharpoons 
\mathrm{C}_2\mathrm{O}_5^{2-}(aq) + 2\mathrm{H}^+
\label{transition3}
\end{equation}
Furthermore, 
We can even see the formation of a tricarbonate ion, C$_3$O$_7^{2-}$(aq) (Fig. \ref{MSM-confinement}b, vii) through the binding of a CO$_2$(aq) molecule with a pyrocarbonate ion under the monolayer nanoconfinement, despite its rarity and inherent instability.

Nanoconfinement shifts the major reaction pathway of the dissolution of CO$_2$(aq) towards reaction (\ref{transition2}), which increases the stability of pyrocarbonate ions in water.  
We analyzed the reaction flux from CO$_2$ to H$_2$CO$_3$ using our MSMs combined with the transition path theory (TPT) \cite{vanden2006towards, metzner2009transition, berezhkovskii2009reactive, noe2009constructing}, and the results are presented in Fig. \ref{MSM-confinement}e. Supplementary Figs. 16, 17, and 18 show the detailed analysis of the flux for each pathway.
In the bulk solution, the formation of H$_2$CO$_3$ predominantly occurs via path I in Fig. \ref{MSM-confinement}d, which involves reaction (\ref{transition1}). As a comparison, under nanoconfinement, the preferred pathway is path II in Fig. \ref{MSM-confinement}d, which is mediated by pyrocarbonate ions, involving the swap of two carbon atoms. Notably, the narrower confinement enhances the formation of 
pyrocarbonate ions, suggesting that adjusting the confinement width can effectively control carbon reactions in aqueous solutions.

The anomalous formation of pyrocarbonate ions in water can be attributed to the extremely spatial confinement. 
First, under the graphene nanoconfinement, the $sp^2$ carbon species tend to align parallel to the solid-liquid interface, and their rotation is hindered \cite{stolte2022nanoconfinement}, which helps to stabilize big carbon species like pyrocarbonate and tricarbonate ions. 
Second, nanoconfinement enhances the breaking of O-H bonds \cite{munoz2017nanoconfinement}, 
so the oxygen atoms in CO$_3^{2-}$ or HCO$_3^{-}$ have more opportunities to form bonds with carbon atoms in CO$_2$(aq).
In fact, we found that under nanoconfinement the diffusion coefficients of protons are about one order of magnitude larger than those of carbon or oxygen atoms, as shown in Supplementary Fig. 19. This stark contrast to the behavior observed in the bulk solution suggests a superionic nature of the confined solutions, which favors the formation of larger oxocarbons.

\subsection {Concerted vs. step-wise proton hopping }

The accurate classification of carbon species in the MSMs enables to identify the transition states of aqueous reactions.
We found that reaction (\ref{transition1}) was primarily facilitated by proton transfer along the hydrogen-bonded water wire (see Fig. \ref{proton}a and Supplementary video for a typical event).
In the bulk solution, reaction (\ref{transition1}) occurred concurrently with the collective proton transfer along the water wire, which usually involved three or four proton jumps, as shown in Fig. \ref{proton}b.  By measuring the time required for three proton jumps, we found that most events occurred within a short period of 200 fs (Fig. \ref{proton}e),  suggesting a concerted behavior. 
The length of the water wire decreased as protons hopped along, similar to the proton transfers reported in the hydroxide and hydronium solutions \cite{marx2006proton, hassanali2011recombination, hassanali2013proton, chen2018hydroxide}. In our study, we did not add \textit{ad-hoc} hydroxide or hydronium ions into the solutions, and the proton hopping was triggered by the carbon species and P-T conditions. 

By performing the same analysis for the nanoconfined solutions, we made two interesting observations. First, we found that in reaction (\ref{transition1}), the water wires under confinement were compressed less than those in the bulk solution, as shown in Fig. \ref{proton}c and d. Second, the proton hopping took a much longer time, showing a stepwise behavior in Fig. \ref{proton}f and g in contrast to that in the bulk solution. These interesting behaviors can also be attributed to the extreme spatial confinement. In the bulk solution, proton hopping typically occurs when a proton vacancy is available, as overcoming a large energy barrier is required otherwise.
Consequently, when the O-H bonds in a connected water wire undergo concerted stretching, the protons hop simultaneously to occupy the vacancies along the water wire, so reaction (\ref{transition1}) can proceed with a reduced energy barrier.
As a comparison, the O-H bonds are easier to break under nanoconfinement. 
Thus, protons are more likely to dissociate from the oxygen atoms even when the subsequent water molecule does not have a vacancy for the incoming proton. As a result, the reaction under nanoconfinement tends to favor a stepwise proton transfer mechanism.

\section{Conclusion}

Our method provides a first-principles framework to construct reaction kinetic networks to systematically elucidate the complex aqueous reactions.
Although here we mainly focused on aqueous carbon reactions, this method can be readily extended to other aqueous reactions.
Unbiased AIMD simulations are so computationally demanding that it is challenging to perform long-time simulations to obtain thermodynamic ensembles of chemical reactions. The MSM 
constructed from many short AIMD simulations with the help of unsupervised machine learning techniques can provide a promising approach to elucidate the reaction kinetics at the quantum mechanics level.
The short AIMD simulations can be carried out in parallel,
enabling the efficient utilization of multiple processing computing.

In this work, with the help of unsupervised machine learning techniques, we constructed MSMs based on AIMD simulations to study the reaction mechanisms and kinetics of dissolved carbon in supercritical water both in the bulk and nanoconfined states. 
It can automatically identify carbon species and reaction pathways without any predefined chemical bonds or reactions.
We found that in the bulk solution, pyrocarbonate ions can easily dissociate. However,
under nanoconfinement, our method revealed their unexpected stability and even the formation of the hypothetical pyrocarbonic acid. 
Nanoconfinement promotes the reaction pathway from CO$_2$(aq) to H$_2$CO$_3$(aq) via the intermediate formation of pyrocarbonate ions, in contrast to the bulk solution where the carbonation reaction pathway is favored, involving proton transfer along a hydrogen-bonded water wire.
The unexpected appearance of pyrocarbonates is related to the superionic behavior of the confined solutions. We also found that the proton transfer in the carbonation reaction is concerted in the bulk solution, but stepwise under nanoconfinement.  
Modulating the 2D confinement can be a highly effective strategy for regulating aqueous chemical reactions.
Our findings suggest that large oxocarbons should not be ignored in aqueous solutions.
Considering that aqueous geofluids are often confined to the nanoscale in pores, grain boundaries, and fractures of Earth's minerals, large oxocarbons may be an important carbon host in the deep carbon cycle, and play an important role in the sequestration of CO$_2$.

\section{Methods}
\subsection{\textit{Ab initio} molecular dynamics}
We performed AIMD simulations using the Qbox code \cite{gygi2008architecture}. The simulations used the Perdew–Burke–Ernzerhof (PBE) exchange-correlation functional \cite{PBE} and the SG15 Optimized Norm-Conserving Vanderbilt (ONCV) pseudopotentials \cite{hamann2013ONCV, schlipf2015ONCV} with a plane-wave kinetic energy cutoff of 65 Ry. When calculating pressure, we increased the cutoff to 85 Ry. For all the simulations, the ratios of C/H/O atomic number were kept unchanged for direct comparisons. Initially, there were 16 CO$_2$ and 32 H$_2$O molecules in the bulk solution, and 12 CO$_2$ and 24 H$_2$O molecules in the bilayer and monolayer solutions under confinement, respectively. All the simulations were performed at a timestep of 0.24 ps in rectangular simulation boxes with periodic boundary conditions. To enable the use of a larger time step, we replaced hydrogen with deuterium but still referred to it as hydrogen. We carried out AIMD simulations in the canonical, i.e., NVT, ensemble. The temperature was controlled by the Bussi-Donadio-Parrinello thermostat \cite{Bussi2007} at a timestep of 24.2 fs.
Our total AIMD simulations exceed 2.68 ns to construct and validate MSMs. More details about the AIMD simulations are provided in Supplementary Methods and Supplementary Tables I.

\subsection {Validation of exchange-correlation functionals}

In our AIMD simulations, we mainly used the semilocal exchange-correlation functional PBE \cite{PBE}. It has been known that generalized gradient approximations may have limitations in accurately describing aqueous solutions under ambient conditions \cite{gillan2016perspective}. However, PBE has demonstrated improved performance under high P and high T conditions, as supported by previous investigations (see, e.g., \cite{Pan2013Dielectric, Pan2014Refractive, pan2016fate}). To validate our results, we conducted a comparative analysis between PBE and the hybrid functional PBE0 \cite{Adamo1999Toward}. The utilization of the PBE0 functional is often favored due to its reduced charge delocalization error, which frequently leads to better agreement with experimental data. 
Our previous study also showed that the concentrations of carbon species predicted by PBE0 are very close to the experimental values \cite{pan2016fate}. 
For an aqueous carbon solution at about 11 GPa and 1000 K, with an initial mole fraction of CO2(aq) of 0.016, both PBE and PBE0 simulations indicate that HCO$_3^-$ predominates as the primary carbon species, with very similar mole percents of 79.8\% and 75.0\%, respectively. Given that neither PBE nor PBE0 functionals adequately capture van der Waals (vdW) interactions, we also investigated the effects of vdW corrections. Specifically, we incorporated Grimme's D3 vdW corrections and Becke-Johnson damping  \cite{Grimme2010Consistent} with the RPBE xc functional (RPBE-D3) \cite{Hammer1999Improved}. Upon studying the solution confined by graphene at $\sim$10 GPa and 1000 K, we observed minimal changes in the mole percentages of carbon species, amounting to less than 6\%. It is worth noting that dispersion interactions do not significantly contribute to the breaking and formation of covalent bonds, and thus, their effects on the investigated chemical speciation are negligible.

\subsection{Procedure of constructing and validating Markov state models}
We constructed MSMs using the MSMbuilder packages \cite{msmbuilder}.
For the bulk solution, we adopted the following procedure.
(i) We built a local coordinate for the oxygen atom O$_j$, which is one of the 
three closest oxygen atoms of the carbon atom C$_i$.
(ii) Using the local coordinates, we extracted the coordinates of the neighboring atoms
of O$_j$, $\{D_{C_iO_j}\}$, as the inputs for principal component analysis (PCA). (iii) We grouped the distribution of PCA into clustered centers using the \textit{k}-means clustering algorithm \cite{likas2003global}. The hyperparameter of the number of clustered centers was chosen by conducting the variational cross-validation with a generalized matrix Rayleigh quotient (GMRQ) \cite{mcgibbon2015variational}, as shown in Supplementary Fig. 2. 
(iv) We assigned a microstate to represent the state of C$_i$ by the union set of the clustered centers of its three closest oxygen atoms.
We ignored the relative order of oxygen atoms.
(v) We lumped the microstates into seven macrostates using the PCCA+ algorithm \cite{deuflhard2005robust}. To better capture the chemical reactions, the number of macrostates is determined based on whether the lumping is able to distinguish the carbon species. 
(vi) We chose the lag time $\tau=1.209$ ps for our MSM based on the implied timescale analysis at the microstate level, as shown in Supplementary Fig. 4. (vii) We further validated our MSM by the Chapman-Kolmogorov (CK) test \cite{prinz2011markov} at the macrostate level (Supplementary Fig. 5).

The construction procedure for nanoconfined solutions is largely the same as that for the bulk solution.
Specifically, we applied the variational cross-validation with GMRQ to determine the hyperparameter of clustering, i.e., the number of clustered centers (Supplementary Figs. 6 and 7), and constructed six- and seven-macrostate-MSMs for the bilayer and monolayer systems, respectively. Note that the clustering can be different (Supplementary Figs. 8 and 9), but our macrostate-MSMs are capable of distinguishing different carbon species, allowing direct comparison of the reaction kinetics among different MSMs. The lag times were chosen to be $\tau=3.628$ ps for two nanoconfined solutions based on their implied timescale analyses (Supplementary Figs. 10 and 11). Then, the MSMs were validated by the CK tests as shown by Supplementary Figs. 12 and 13 at the macrostate level. 

\subsection{Flux analysis and mean first passage time}
The fluxes of the reaction pathways were analyzed in the mathematical framework of the transition path theory \cite{vanden2006towards, metzner2009transition, berezhkovskii2009reactive, noe2009constructing}. We computed the ensembled fluxes of reactions using the transition probability matrix at the microstate level. The net fluxes of transitions were then assembled into the coarse-grained macrostates. We also performed Markov Chain Monte Carlo (MCMC) simulations to generate independent trajectories based on the transition probability matrix from the microstate-MSM. The generated MCMC trajectories were utilized to calculate the MFPT between each pair of macrostates (see Supplementary Methods for more detail).

\section{Data Availability}

The data that support this study are available upon request from the authors.

\section{Code Availability}

Qbox and MSMbuilder are free and open-source codes available at http://qboxcode.org and http://msmbuilder.org. Data processing scripts and codes are available upon request from the authors.

\section{Acknowledgements}
We thank X. Huang for many useful discussions. Y.Y. gratefully acknowledges the National Natural Science Foundation of China/Research Grants Council Joint Research Scheme Grant N\_HKUST635/20 and Hong Kong Research Grant Council (HKRGC) Grant 16308321. 
D.P. acknowledges support from the Croucher Foundation through the Croucher Innovation Award, Hong Kong Research Grants Council (Projects GRF-16306621, GRF-16301723, and C1006-20WF), National Natural Science Foundation of China through the Excellent Young Scientists Fund (22022310), and the Hetao Shenzhen/Hong Kong Innovation and Technology Cooperation (HZQB-KCZYB-2020083). 
Part of this work was carried out using computational resources from the National Supercomputer Center in Guangzhou, China, and the X-GPU cluster supported by the HKRGC Collaborative Research Fund C6021-19EF.

\section{Competing Interests}
The authors declare no competing interests.

\bibliography{ref}

\newpage

\begin{figure}
\centering
\vspace{5mm}
\includegraphics[width=0.7 \textwidth]{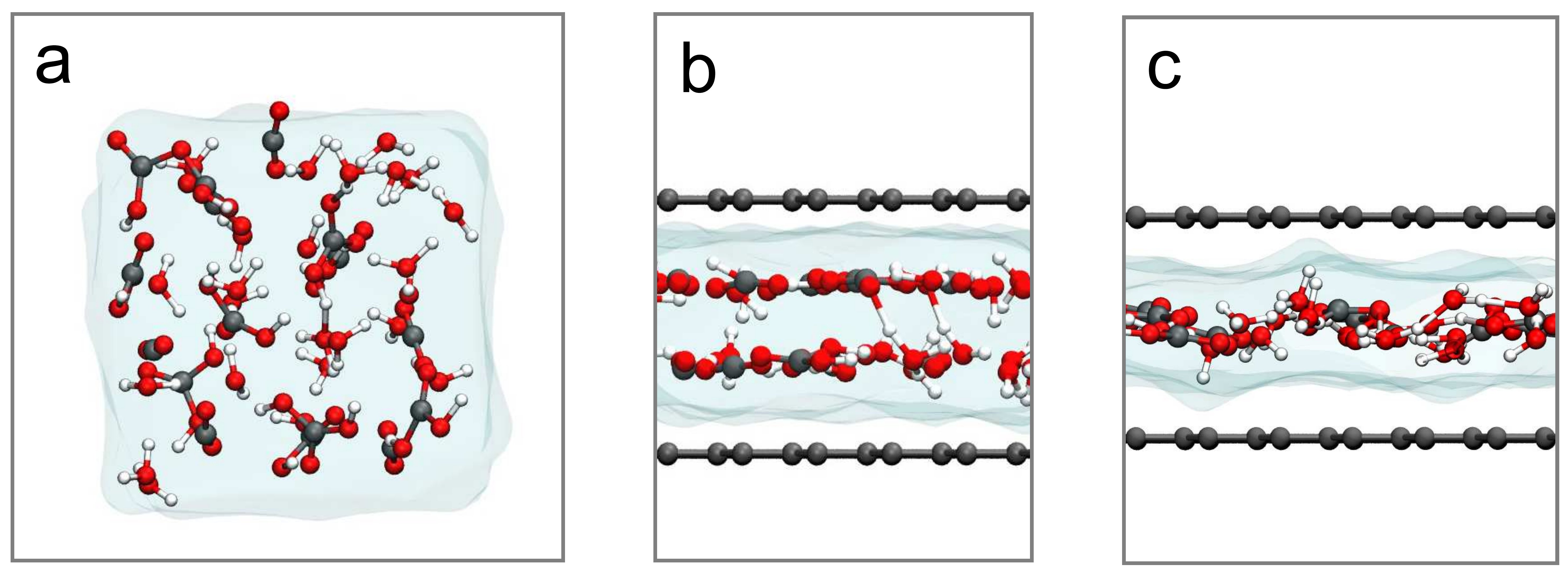}
\caption{Snapshots of dissolved carbon in supercritical water at $\sim$ 10 GPa and 1000 K. The bulk solution is shown in (\textbf{a}).
The confinement widths, i.e. the distances between two model graphene sheets, are 0.775 and 0.650 nm, resulting in the bilayer (\textbf{b}) and monolayer (\textbf{c}) solutions, respectively. 
Carbon, oxygen, and hydrogen atoms are represented by gray, red, and white spheres, respectively.
The three snapshots were randomly selected from \textit{ab initio} molecular dynamics simulations trajectories.
}
\label{system}
\end{figure}

\begin{figure}
\centering
\vspace{5mm}
\includegraphics[width=1 \textwidth]{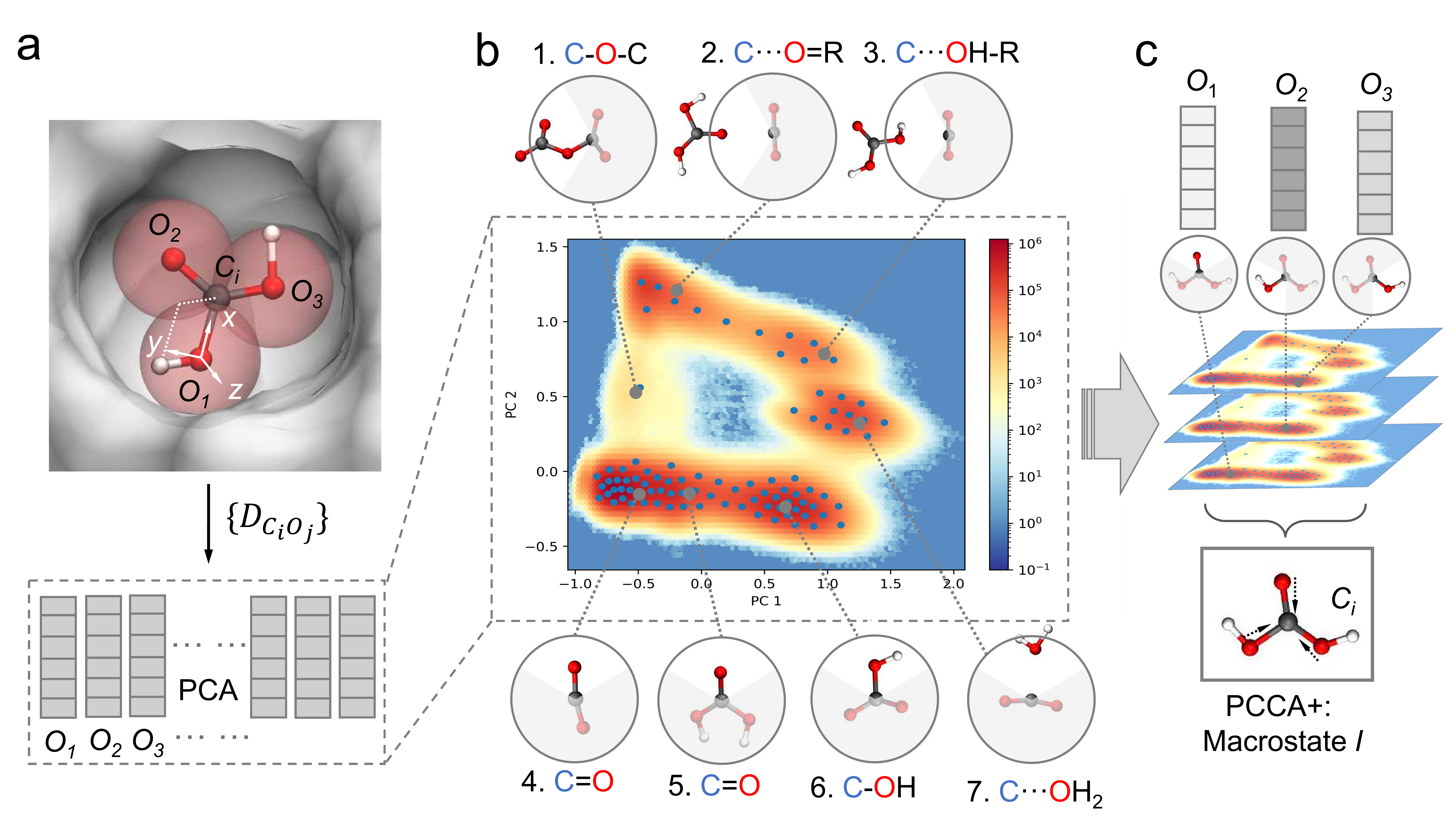}
\caption{Procedure of constructing Markov state models. \textbf{a}. Principal component analysis (PCA) using the local atomic coordinates, \{$D_{C_iO_j}$\}, around the oxygen atom O$_j$, which is one of the three oxygen atoms neighboring to the carbon atom C$_i$. \textbf{b}. Distribution of PCA on the top two PCs.
The distribution detects the chemical environment of O$_j$ and its relation with C$_i$.  
In each representative region 
(\textbf{1-7}), C$_i$ (blue text) is located at the center of the circle with O$_j$ (red text) and its surroundings (black text, with no shade in the circle).
\textbf{c}. Assignment of microstates for C$_i$ by combining the clustered centers of the three neighboring oxygen atoms of C$_i$. The PCCA+ algorithm was used to group the microstates into the macrostates.}
\label{framework}
\end{figure}

\begin{figure}
\centering
\vspace{5mm}
\includegraphics[width=1.0 \textwidth]{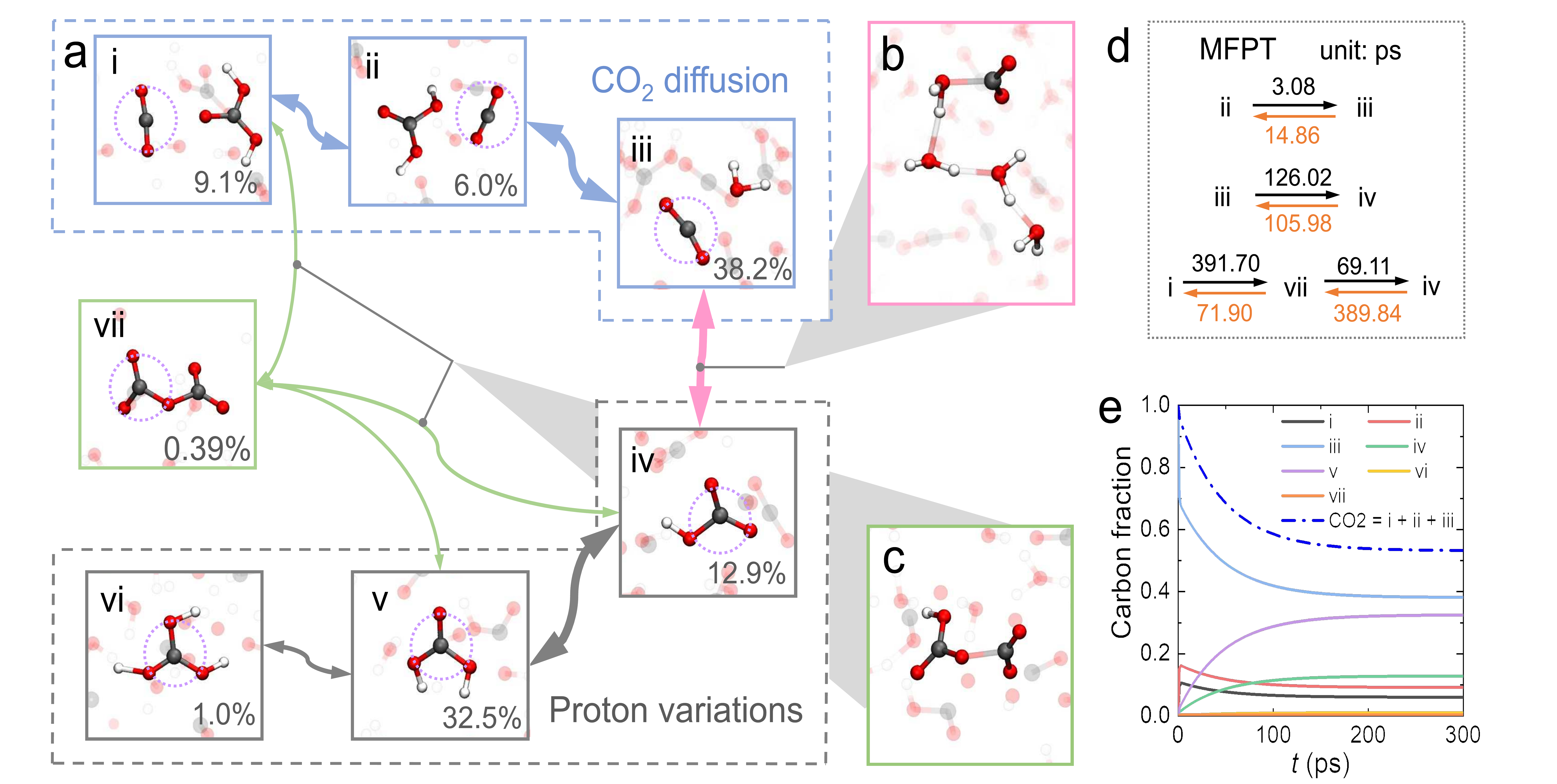}
\caption{Reaction kinetic network of dissolved carbon in the bulk solution at $\sim$ 10 GPa and 1000 K obtained by the Markov state model (MSM). \textbf{a}. Representative snapshots of the seven macrostates (i-vii) from the MSM. Each macrostate represents a state of the central carbon atom (labelled by a purple dashed circle), which assumes a carbon species or is surrounded by different molecules.  
The atoms in the background are blurred for better visualization. 
The bidirectional arrows indicate the major state transitions. 
The stationary population (i.e., carbon mole percent) is shown for each state.
\textbf{b}. The transition state from state iii to state iv obtained by retracing the state change in the original trajectories. It shows the hydrogen-bonded water wire mediates reaction (\ref{transition1}). 
\textbf{c}. The transition state in reaction (\ref{transition2}) for the formation of pyrocarbonate ions. 
\textbf{d}. Mean first passage times (MFPTs) for the diffusion of CO$_2$ between states ii and iii, and the key transitions in (\textbf{b}) and (\textbf{c}). 
\textbf{e}. Time evolution of carbon species obtained from the MSM. The calculation started from state iii.}
\label{MSM-bulk}
\end{figure}

\makeatletter

\newenvironment{Figure}{%
\par\addvspace{12pt plus2pt}%
\def\@captype{figure}%
}{%
\par\addvspace{12pt plus2pt}%
}%
\long\def\@makecaption#1#2{%
\vskip\abovecaptionskip
\sbox\@tempboxa{#1. #2}%
\ifdim \wd\@tempboxa >\hsize
#1. #2\par
\else
\global \@minipagefalse
\hb@xt@\hsize{\hfil\box\@tempboxa\hfil}%
\fi
\vskip\belowcaptionskip}

\makeatother

\begin{Figure}
\centering
\vspace{5mm}
\includegraphics[width=1.0 \textwidth]{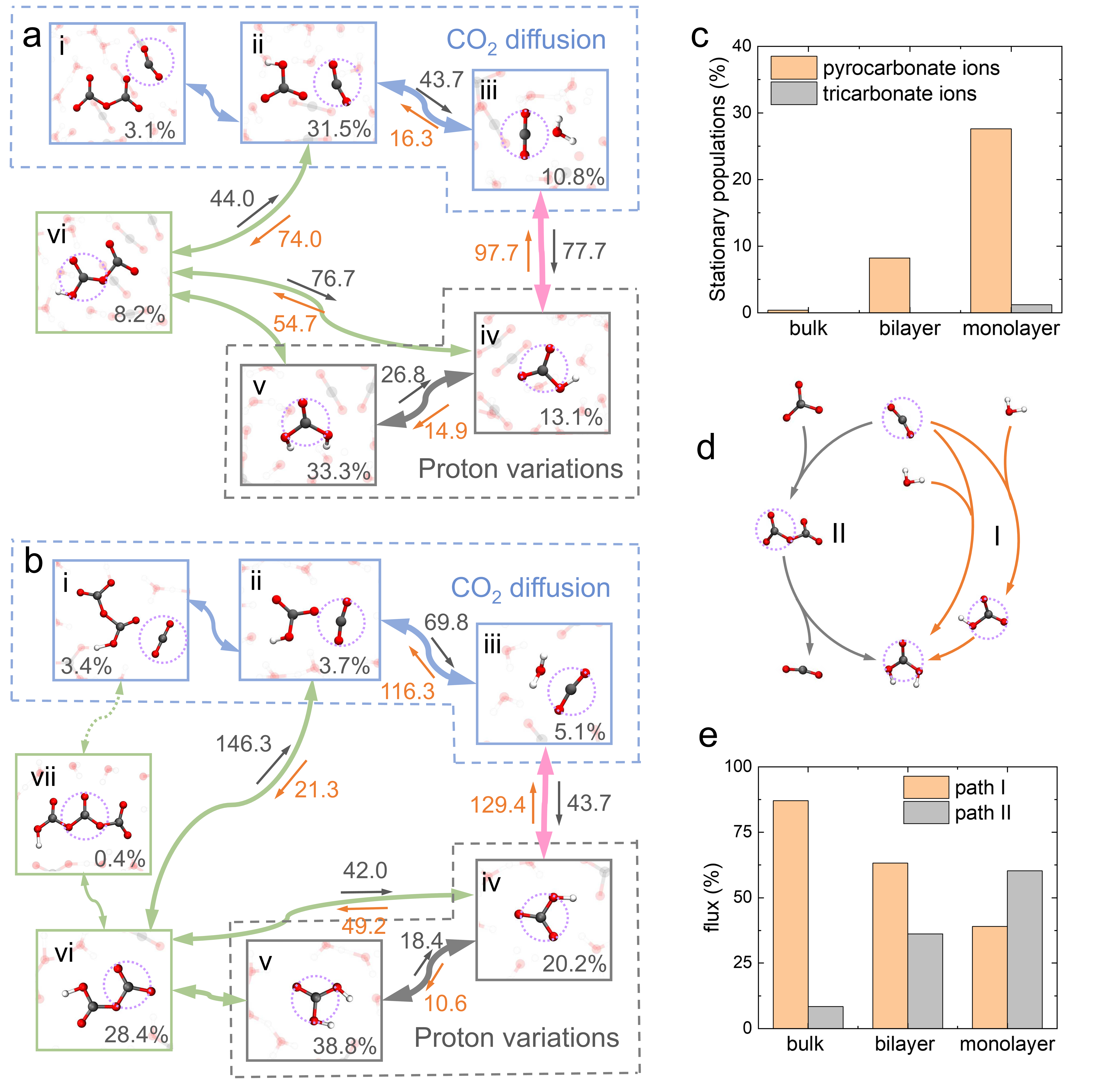}
\caption{Reaction kinetic networks of aqueous carbon under nanoconfinement at $\sim$ 10 GPa and 1000 K. The bilayer (\textbf{a}) and monolayer (\textbf{b}) confinements are compared. The central carbon atom in each state is labelled by a purple dashed circle.
The stationary population (i.e., carbon mole percent) is shown for each state.
The key mean first passage times (MFPTs, unit: ps) are given along the reaction arrows.
\textbf{c}. Stationary polulations (i.e., mole percents) of pyrocarbonate and tricarbonate ions in the three solutions.
\textbf{d}. Illustration of two typical reaction pathways, I and II, from CO$_2$(aq) to H$_2$CO$_3$(aq).
\textbf{e}. Fluxes of pathways I and II in the three solutions.  
}
\label{MSM-confinement}
\end{Figure}

\newpage
\begin{Figure}
\centering
\vspace{5mm}
\includegraphics[width=1 \textwidth]{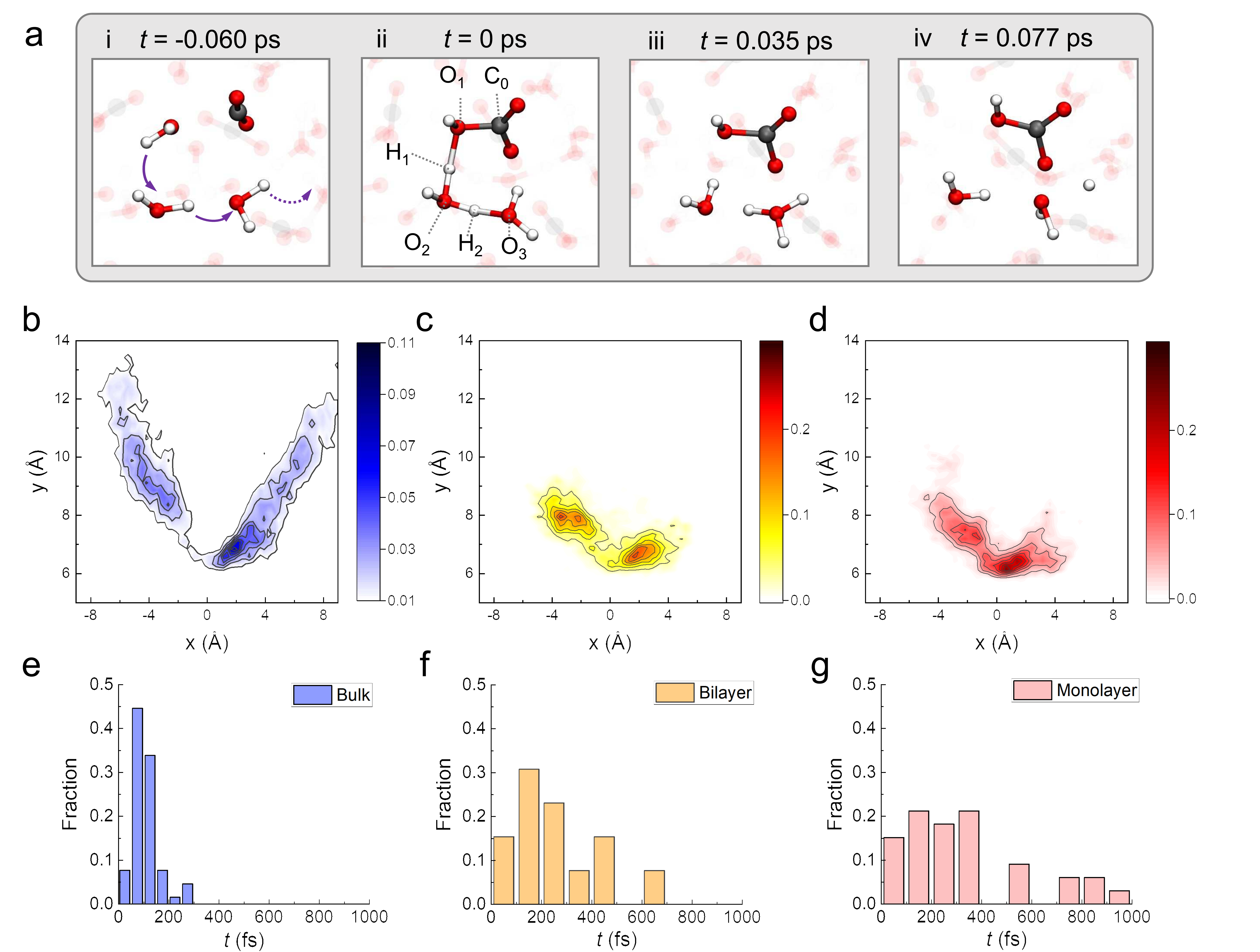}
\caption{
Proton transfer in the aqueous carbon reactions.
\textbf{a}. Time-lapse snapshots of the proton transfer consisting of double or triple proton jumps, labelled by the purple arrows in (i), in the bulk solution.
The water wire made by C$_0-$O$_1-$O$_2-$O$_3$ was formed spontaneously in (ii). 
The density maps show the probability distributions of proton positions and water wire lengths in the bulk (\textbf{b}), bilayer (\textbf{c}), and monolayer (\textbf{d}) solutions.
The $x$ coordinate describes the relative displacement of protons in the two connected proton jumps: $x=|r_{H1}-r_{O1}|-|r_{H1}-r_{O2}|+|r_{H2}-r_{O2}|-|r_{H2}-r_{O3}|$, where $r_i$ denotes the position of the atom $i$.
The $y$ coordinate is the water wire length, $y=|r_{C0}-r_{O1}|+|r_{O1}-r_{O2}|+|r_{O2}-r_{O3}|$. 
\textbf{e}-\textbf{g}. 
Distribution of proton transfer time for three proton jumps in the three solutions. }
\label{proton}
\end{Figure}

\end{document}


\title{Supplementary Information for unveiling the hidden reaction kinetic network of carbon dioxide in supercritical aqueous solutions
}
\author{Chu Li}
\email{cliag@connect.ust.hk}
\affiliation{Department of Physics, The Hong Kong University of Science and Technology, Hong Kong, China}
\author{Yuan Yao}
\affiliation{Department of Mathematics, The Hong Kong University of Science and Technology, Hong Kong, China}
\author{Ding Pan}
\email{dingpan@ust.hk}
\affiliation{Department of Physics, The Hong Kong University of Science and Technology, Hong Kong, China}
\affiliation{ Department of Chemistry, The Hong Kong University of Science and Technology, Hong Kong, China}
\affiliation{HKUST Shenzhen-Hong Kong Collaborative Innovation Research Institute, Shenzhen, China}

\maketitle

\section{Supplementary Methods}
\subsection{1. Confinement potential}
We confined the solutions by using two graphene sheets perpendicular to the $Z$ direction with a distance of $h_Z$. A vacuum space of 7 {\AA} is maintained, so the solutions are at least 10 {\AA} away from their replicas with periodic boundary conditions.
We put the graphene-H$_2$O forces on oxygen atoms, 
which were fitted to the diffusion Monte Carlo calculations for the two-legged water molecules adsorbed on the graphene sheet [37]. 
The interaction potential as a function of the perpendicular distance $d$ between graphene and atoms has a Morse shape of 
\begin{equation}
E^{\mathrm{O}}(d) = D^{\mathrm{O}} \left[ \left( 1 - e^{-a^{\mathrm{O}}(d-d_0^{\mathrm{O}})}  \right)^2 -1 \right],
\end{equation}
where $D^{\mathrm{O}} = 9.55185$ kJ mol$^{-1}$, $a^{\mathrm{O}} = 1.34725$ {\AA}$^{-1}$, and $d_0^{\mathrm{O}} = 3.37265$ {\AA}.

We fitted the graphene-CO$_2$ interactions calculated from the optB86-vdW functional [38] by using the following Morse potential:
\begin{equation} \label{g-CO2}
E^{\mathrm{CO_2}}(d) = D^{\mathrm{CO_2}} \left[ \left( 1 - e^{-a^{\mathrm{CO_2}}(d-d_0^{\mathrm{CO_2}})}  \right)^2 -1 \right],
\end{equation} 
where $D^{\mathrm{CO_2}} = 21.50492$ kJ mol$^{-1}$, $a^{\mathrm{CO_2}} = 1.18449$ {\AA}$^{-1}$, and $d_0^{\mathrm{CO_2}} = 3.25282$ {\AA}. 
Then the graphene-carbon interaction ($E^{\mathrm{C}}(d)$) was obtained by subtracting the graphene-oxygen interaction and adding a short-distance repulsive term to avoid nonphysical interaction at $d<3$ \AA. 
\begin{equation} \label{g-C} 
E^{\mathrm{C}}(d) = E^{\mathrm{CO_2}}(d) - 2E^{\mathrm{O}}(d) + D^{\mathrm{rep}} e^{-2 a^{\mathrm{rep}} (d-d_0^{\mathrm{rep}})}, 
\end{equation}  
where $D^{\mathrm{rep}} = 18.69565$ kJ mol$^{-1}$, $a^{\mathrm{rep}} = 26.09666$ {\AA}$^{-1}$, and $d_0^{\mathrm{rep}} = 2.56637$ {\AA}.

For direct comparison, we performed all the \textit{ab initio} molecular dynamics (AIMD) simulations with the same initial CO$_2$(aq) concentration and under the same $P-T$ conditions (i.e., $P=10$ GPa, $T= 1000$ K, see Supplementary Table I.). 
We used the lateral pressure ($P_{\parallel}$) for the confined solutions: $P_{\parallel}=\frac{\sigma_{xx}^{'}+\sigma_{yy}^{'}}{2}$, where $\sigma_{kk}^{'}=\sigma_{kk}(l_Z/h_Z)$ represents the modified diagonal element of the stress tensor $\sigma_{kk}$, accounting for the presence of vacuum space in the unit cells [13, 39, 41].


\subsection{2. Constructing and validating Markov State Models}


The concept of MSM is to model the continuous dynamics as discrete Markovian jumps in the discretized partitions of the relevant space at a time interval, i.e., lag time. For the reactions of carbon in supercritical water, we focused on tracing the reactions associated with carbon. Thus, from the AIMD trajectories, we extracted and clustered the reaction space of carbon into a discrete number of states, and computed the transition probability between each pair of states at a discrete time interval (lag time $\tau$). If $\tau$ is appropriate to allow the fast motions to be integrated out, the MSM is Markovian, which means the probability of a state that a carbon atom is going to assume at $t+\tau$ only depends on its current state at $t$. Under the Markovian assumption, the reaction dynamics can be modeled by a sequence of discrete Markov jumps in the reaction space. And the long-time dynamics can be modeled by the first-order master equation: 
\begin{equation}
    P(n\tau)=P(0)T^n(\tau),
\label{timepropagation}
\end{equation}
where $P(n\tau)$ represents the probability vector of microstates assumed by the central carbon atom at time $n\tau$. $T(\tau)$ is the transition probability matrix with a lag time $\tau$.


\subsection{2.1. Extraction and partition of the reaction space of carbon}
From AIMD simulations, we obtained the trajectory of each carbon atom together with its surrounding atomic environment. Specifically, for the carbon atom $C_i$, we established three local coordinate systems, originating from its three nearest oxygen atoms ($O_j$, where $j=1,2$, or $3$), as shown in Fig. 1 (a) in the main text. The local coordinates were built according to the following rules: (1) The position of $O_j$ is defined as the origin; (2) $x$-axis is oriented along $\overrightarrow{O_jC_i}$; (3) Perpendicular to the $x$-axis, $y$-axis is located at the $C_iO_jH$ plane, where $H$ is the closest hydrogen atom of $O_j$. (4) $z$-axis is set perpendicular to the $xy$ plane based on the right-hand rule. The atomic environment encoded in the local coordinates is extracted as input:  ${D_{C_iO_j}}=\{\frac{1}{r_{O_jX}^2}, \frac{x_{O_jX}}{r_{O_jX}^3},  \frac{y_{O_jX}} {r_{O_jX}^3}, \frac{z_{O_jX}}{r_{O_jX}^3}\}$, where $X$ denotes the neighboring atoms of $O_j$ including two carbon atoms, five oxygen atoms, and eight hydrogen atoms, resulting in a $(2+5+8)*4=60$ dimensional vector. Here, $r,x_{O_jX},y_{O_jX}$, and $z_{O_jX}$ are the length, $x, y$, and $z$ components of the vector $\overrightarrow{O_jX}$, respectively. In $D_{C_iO_j}$, $X$ is first sorted based on the element type, and then $r$.

\begin{figure}[H]
\centering
\vspace{5mm}
\includegraphics[width=0.5 \textwidth]{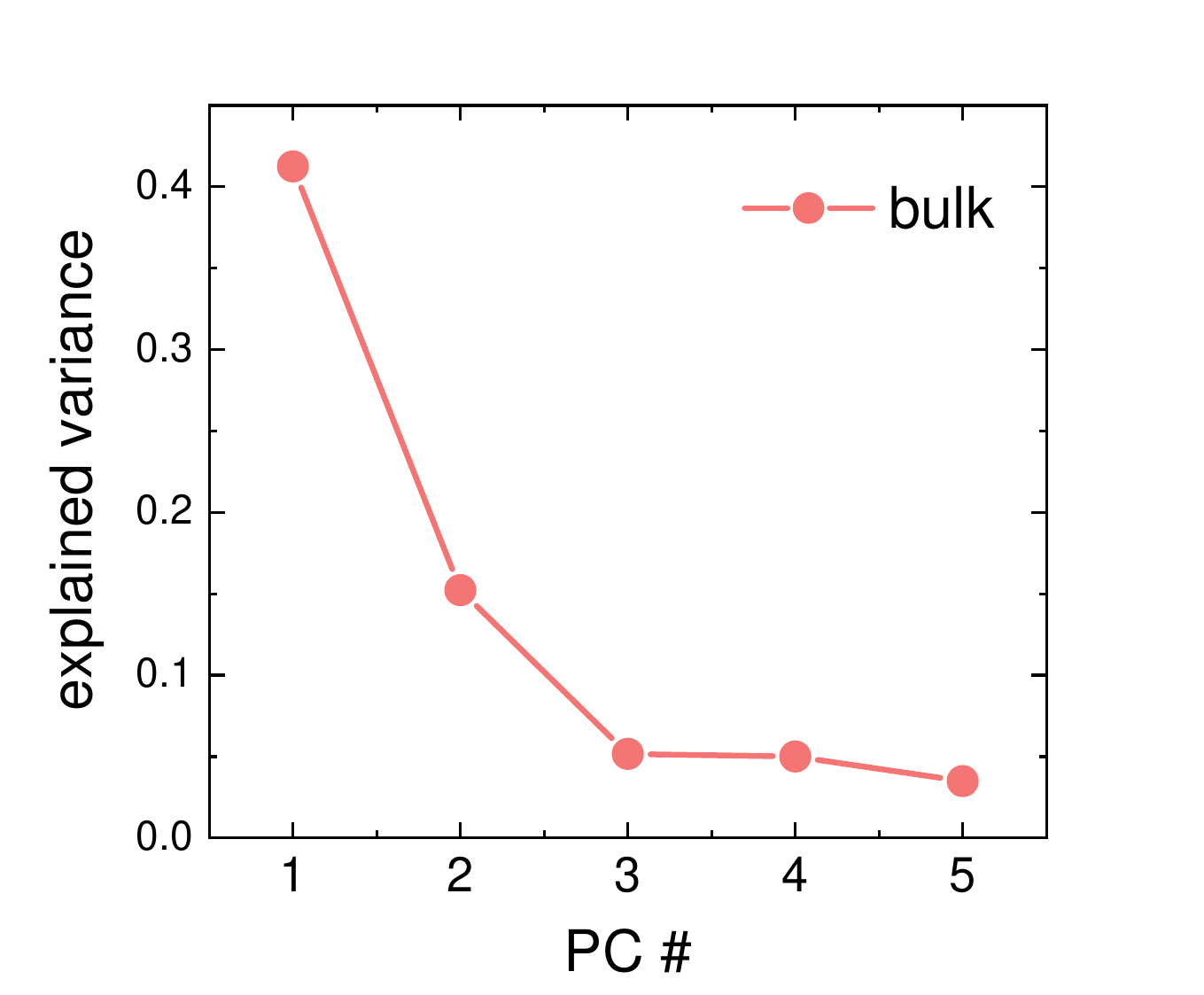}
\caption{Explained variance of the principal component analysis for the bulk solution. Explained variance refers to the amount of variance that can be attributed to each individual principal component (PC). }
\label{PC-bulk}
\end{figure}

All the inputs $\{D_{C_iO_j}\}$ were collected orderlessly for PCA. Supplementary Fig. \ref{PC-bulk} shows 
the fractions of variance explained by the top five principal components. 
We maintained the top two PCs because these two PCs are able to capture the most information, which effectively reduces the dimensionality from 60 to 2. 
The number of the clustered centers ($k=90$) was selected based on the variational cross-validation of GMRQ [66], as shown in Supplementary Fig. \ref{GMRQ-bulk}. To improve connectivity, we further separately merged the clustered centers with high density in ($\mathbf{4}$), ($\mathbf{5}$), and ($\mathbf{6}$) in Fig. 2b in the main text. 
Then we created a microstate to represent C$_i$ as the union set of three cluster centers of its neighboring O$_j$.


\subsection{2.2. Validations of the MSM for the bulk solution}
We lumped microstates into macrostates using the PCCA+ algorithm [45] with a time interval of 0.242 ps to better visualize the chemical reactions. Supplementary Fig. \ref{carbon-bulk} shows our macrostate lumping can precisely detect different carbon species and their environment. We then selected the lag time $\tau=1.209$ ps and validated our model by implied timescale analysis [23] at the microstate level in Supplementary Fig. \ref{ITS-bulk}. We also validated our MSM using the Chapman-Kolmogorov (CK) test [23] at the macrostate level in Supplementary Fig. \ref{CKtest-bulk}. Eventually, we obtained a 7-macrostate-MSM as shown in Fig. 3 in the main text. 

\subsection{2.3. Validations of the MSMs for the nanoconfined solutions}
Supplementary Figs. \ref{PCA-bilayer} and \ref{PCA-monolayer} present the PCA on the top two PCs for the bilayer and monolayer solutions, respectively. The hyperparameters, i.e. number of centers ($k$), were chosen by performing variational cross-validation with GMRQ, as shown in Supplementary Figs. \ref{GMRQ-bi} and \ref{GMRQ-mo}, respectively. We lumped microstates into macrostates using PCCA+ with a time interval of 0.363 and 0.484 ps, and obtained 6- and 7-macrostates MSMs for the bilayer and monolayer solutions.
The lag times $\tau=3.628$ ps were selected for both bilayer and monolayer solutions based on their implied timescale analyses at the microstate level (Supplementary Figs. \ref{ITS-bilayer} and \ref{ITS-monolayer}). Then we validated the MSMs by the CK tests in Supplementary Figs. \ref{CKtest-bilayer} and \ref{CKtest-monolayer}.  

\subsection{3. Markov Chain Monte Carlo simulations and mean first passage time}
By applying the transition probability matrix $T(\tau)$, we conducted Markov Chain Monte Carlo (MCMC) simulations to obtain kinetic as Markov jumps between microstates. We generated long time trajectories using the following relation:
\begin{equation}
P(t+\tau)=P(t)T(\tau)
\label{propagation}
\end{equation}
where $P(t)$ is the probability of microstates at time $t$. After obtaining long time MCMC trajectories and transforming them into macrostate trajectories, we calculated the mean first passage time (MFPT) for each transition between two macrostates by averaging their first transition time.

\subsection{4. Flux analysis and time dynamics of carbon species}
We analyzed the fluxes of reaction pathways based on the
transition path theory (TPT) [46-49]. We computed the fluxes along with their relative probabilities at the microstate level, 
and assembled the net fluxes 
into coarse-grained transitions among macrostates.

The time evolution of carbon species can be obtained by propagating the transition probability matrix according to Eq. (\ref{timepropagation}) at the microstate level. 
The total probability of all the microstates was initialized to 1.
By assembling $P(n\tau)$ into macrostates, we can compute the temporal dynamics of carbon species. 
Fig. 3e in the main text shows an example of the time evolution starting from an evenly-distributed macrostate labeled as ``iii''.
This is achieved by setting
$P_{i_{iii}}(0)$ to $\frac{1}{n}$,
where $i_{iii}$ represents microstate i belonging to macrostate iii, and $n$ is the number of microstates in state iii.

\newpage

\section*{Supplementary Tables}
\begin{table}[H]
\caption{Details of the AIMD simulations for three solutions.  $P$ is the bulk pressure or the lateral pressure ($P_{\parallel}$) under nanoconfinement. $T$ is the temperature of simulations. n(CO$_2$) and n(H$_2$O) are the initial numbers of CO$_2$ and H$_2$O in the simulation boxes. $l_x$, $l_y$, and $l_z$ are the dimensions of simulation boxes. $h_z$ is the distance between two confining graphene sheets.  $t$ is the simulation time. }
\label{AIMD}
\centerline{
\begin{tabular}{ c  c   c  c  c  c  c  c  c   }
\hline 
\hline
solution$\quad$  &  $\quad$ $P$(GPa)$\quad$      &  $T$(K)   & $\quad$ n(CO$_2$) $\quad$& n(H$_2$O)$\quad$ &  $l_x=l_y$ (nm)  & $\quad$ $l_z$ (nm) $\quad$&   $\quad$ $h_z$ (nm) &$\quad$ $t$(ps)$\quad$\\
\hline
Bulk $\quad$ & 9.3$\pm$1.2              & 1000 & 16  & 32 & 1.049  & 1.049  & $-$ & 1287\\
\hline
Bilayer$\quad$ & 10.94$\pm$1.36              &  1000  & 12  & 24 & 1.236  & 1.475  & 0.775  & 857\\
\hline
Monolayer$\quad$& 9.35 $\pm$0.55  & 1000  &  12 & 24  & 1.550 & 1.350  & 0.650 & 549\\
\hline
\hline
\end{tabular}
}
\end{table}
\newpage
\begin{table}[H]
\caption{Mean first passage time (MFPT, unit: ps) of the transition from a state in the row to a state in the column for the bulk solution in Fig. 3. The MFPTs were computed from Markov Chain Monte Carlo (MCMC) trajectories. The uncertainty is the standard error obtained from five independent long trajectories. }
\label{MFPT-bulk}
\centerline{
\begin{tabular}{ c  c   c  c  c  c  c  c  c }
\hline 
\hline
State  & $\qquad$i $\qquad$     & $\qquad$ ii$\qquad\qquad$   & $\qquad$ iii$\qquad$ & $\qquad$ iv$\qquad \qquad$ & $\qquad$ v$\qquad $  & $\qquad$ vi$\qquad\qquad$ & $\qquad$ vii$\qquad $\\
\hline
i$\quad$ & $-$              & 13.96 $\pm$ 0.09 & 3.14$\pm$ 0.04  & 126.54$\pm$ 0.97 & 124.65$\pm$ 1.06  & 244.10$\pm$ 2.04  & 391.70$\pm$ 2.62\\
\hline
ii$\quad$ & 19.76$\pm$0.15              &  $-$  & 3.08$\pm$ 0.03  & 126.48$\pm$ 0.92 & 124.70$\pm$ 1.07  & 244.04$\pm$ 2.32  & 391.61$\pm$ 2.80 \\
\hline
iii $\quad$& 20.71 $\pm$0.18    & 14.86 $\pm$ 0.14   &  $-$& 126.02 $\pm$ 0.93    & 124.20 $\pm$ 1.08 & 243.63 $\pm$ 2.33  & 392.16 $\pm$ 3.23\\
\hline
iv $\quad$ & 123.94 $\pm$ 1.01 &118.48 $\pm$0.98 & 105.98 $\pm$0.77 &$-$& 5.21 $\pm$ 0.09 & 123.75$\pm$1.27 &  389.84 $\pm$ 4.58      \\
\hline
v $\quad$ &  125.03 $\pm$ 1.12 & 119.59 $\pm$ 0.90 & 107.05 $\pm$ 0.82 & 8.27 $\pm$ 0.09 &$-$ & 122.18 $\pm$ 1.18 & 390.77 $\pm$ 4.52\\
\hline
vi $\quad$ & 125.37 $\pm$ 0.98 & 120.00 $\pm$ 1.00 & 107.48 $\pm$ 0.71 & 8.22 $\pm$ 0.13 & 3.58 $\pm$ 0.10 & $-$ & 390.48 $\pm$ 4.91\\
\hline
vii $\quad$ & 71.90 $\pm$ 1.02 & 66.03 $\pm$ 1.07 & 54.57 $\pm$ 0.99 & 69.11 $\pm$ 0.98 & 66.60 $\pm$ 1.14 & 184.35 $\pm$ 2.72 &$-$
\\
\hline
\hline
\end{tabular}
}
\end{table}

\newpage
\begin{table}[H]
\caption{Mean first passage time (MFPTs, unit: ps) of the transition from a state in the row to a state in the column for the bilayer solution in Fig. 4a. The uncertainty is the standard error obtained from five independent MCMC trajectories. }
\label{MFPT-bilayer}
\centerline{
\begin{tabular}{ c  c   c  c  c  c  c  c  }
\hline 
\hline
State  & $\qquad$i $\qquad$     & $\qquad$ ii$\qquad\qquad$   & $\qquad$ iii$\qquad$ & $\qquad$ iv$\qquad \qquad$ & $\qquad$ v$\qquad $  & $\qquad$ vi$\qquad\qquad$ \\
\hline
i$\quad$ & $-$              & 19.21 $\pm$ 0.12 & 51.19$\pm$ 0.27  & 73.88$\pm$ 0.36 & 67.52$\pm$ 0.64  & 65.55$\pm$ 0.29  \\
\hline
ii$\quad$ & 123.45$\pm$0.61              &  $-$  & 43.72$\pm$ 0.21  & 79.03$\pm$ 0.31 & 72.19$\pm$ 0.42  & 74.00$\pm$ 0.15  \\
\hline
iii $\quad$& 127.76 $\pm$0.61    & 16.29 $\pm$ 0.09   &  $-$& 77.66 $\pm$ 0.26    & 70.33 $\pm$ 0.38 & 76.09 $\pm$ 0.35  \\
\hline
iv $\quad$ & 170.74 $\pm$ 1.13 & 71.38 $\pm$0.45 & 97.72 $\pm$0.71 &$-$& 14.93 $\pm$ 0.09 & 76.70$\pm$0.42  \\
\hline
v $\quad$ &  174.25 $\pm$ 1.10 & 74.18 $\pm$ 0.46 & 100.21 $\pm$ 0.70 & 26.83 $\pm$ 0.09 &$-$ & 80.04 $\pm$ 0.44 \\
\hline
vi $\quad$ & 139.89 $\pm$ 0.84 & 44.04 $\pm$ 0.19 & 73.87 $\pm$ 0.22 & 54.67 $\pm$ 0.27 & 48.08 $\pm$ 0.30 & $-$ \\

\hline
\hline
\end{tabular}
}
\end{table}

\newpage
\begin{table}[H]
\caption{Mean first passage time (MFPT, unit: ps) of the transition from a state in the row to a state in the column for the monolayer solution in Fig. 4b. The uncertainty is the standard error obtained from five independent MCMC trajectories. }
\label{MFPT-monolayer}
\centerline{
\begin{tabular}{ c  c   c  c  c  c  c  c  c }
\hline 
\hline
State  & $\qquad$i $\qquad$     & $\qquad$ ii$\qquad\qquad$   & $\qquad$ iii$\qquad$ & $\qquad$ iv$\qquad \qquad$ & $\qquad$ v$\qquad $  & $\qquad$ vi$\qquad\qquad$ & $\qquad$ vii$\qquad $\\
\hline
i$\quad$ & $-$              & 111.99 $\pm$ 0.60 & 71.26 $\pm$ 0.09  & 46.76$\pm$ 0.19 & 39.83$\pm$ 0.15  & 19.99$\pm$ 0.12  & 1534.69$\pm$ 18.66\\
\hline
ii$\quad$ & 134.84 $\pm$1.39          &  $-$  & 69.79$\pm$ 0.35  & 45.62 $\pm$ 0.15 & 38.54$\pm$ 0.17  & 21.36$\pm$ 0.14  & 1535.56$\pm$ 16.93 \\
\hline
iii $\quad$& 139.87 $\pm$1.13    & 116.35 $\pm$ 0.23  &  $-$& 43.76 $\pm$ 0.11  &  36.64$\pm$ 0.15 & 21.82 $\pm$ 0.09  & 1539.79 $\pm$ 19.86\\
\hline
iv $\quad$ & 202.43 $\pm$ 1.16 &178.49 $\pm$0.43 & 129.42 $\pm$0.22 &$-$& 10.63 $\pm$ 0.02 & 49.20$\pm$0.08 &  1559.74 $\pm$ 15.52      \\
\hline
v $\quad$ &  203.65 $\pm$ 1.30 & 179.39 $\pm$ 0.54 & 130.08 $\pm$ 0.22 & 18.41 $\pm$ 0.04 &$-$ & 50.30 $\pm$ 0.13 & 1559.68 $\pm$ 17.09\\
\hline
vi $\quad$ & 168.32 $\pm$ 1.12 & 146.31 $\pm$ 0.33 & 100.40 $\pm$ 0.22 & 41.99 $\pm$ 0.14 & 35.28 $\pm$ 0.12 & $-$ & 1522.27 $\pm$ 16.74\\
\hline
vii $\quad$ & 172.03 $\pm$ 1.00 & 149.66 $\pm$ 1.23 & 105.29 $\pm$ 1.27 & 46.67 $\pm$ 0.68 & 39.13 $\pm$ 0.44 & 13.43 $\pm$ 0.14 &$-$
\\

\hline
\hline
\end{tabular}
}
\end{table}

\newpage

\section*{Supplementary Figures}

\newpage
\begin{figure}[H]
\centering
\vspace{5mm}
\includegraphics[width=0.5 \textwidth]{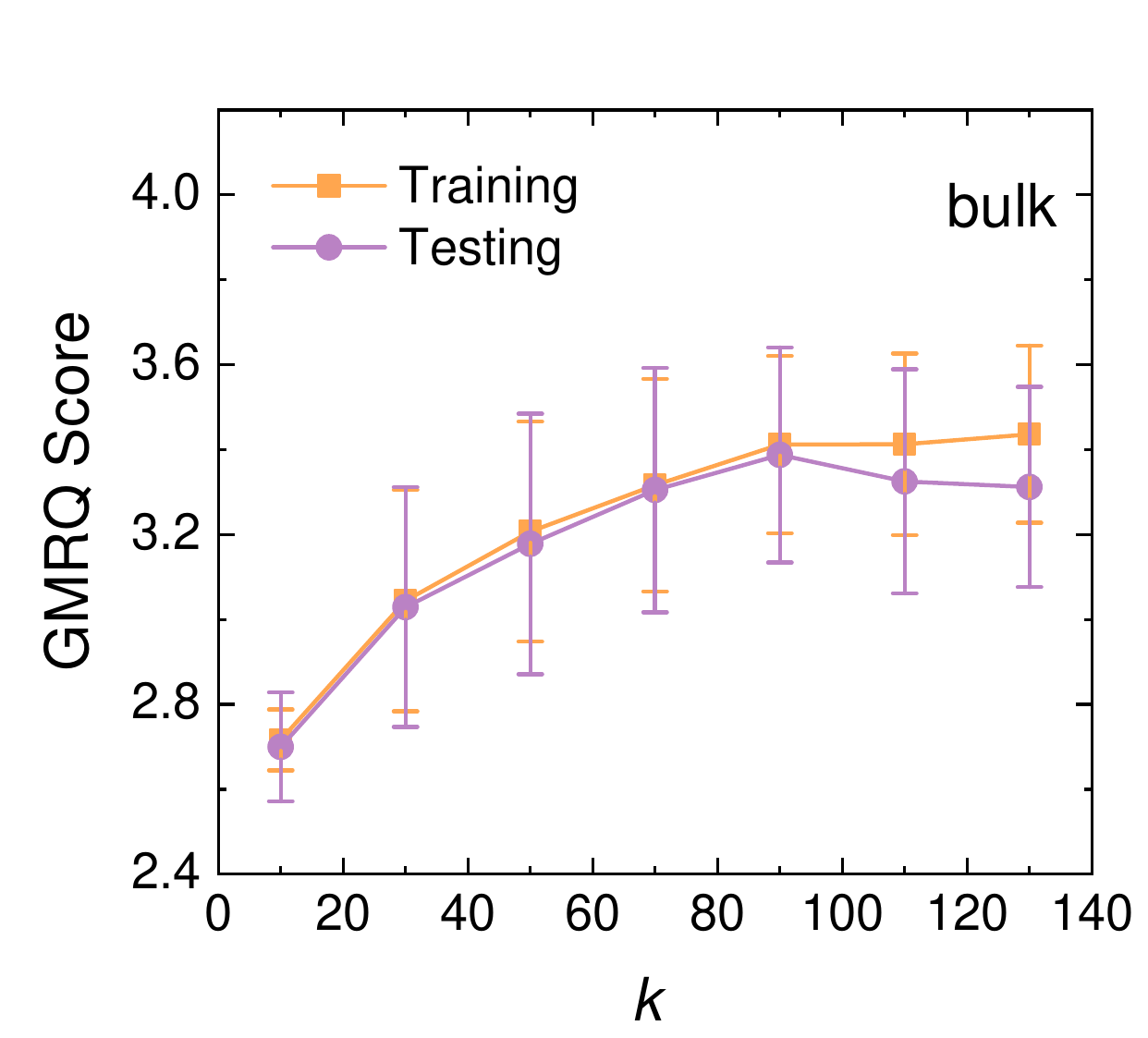}
\caption{The number of centers for the $k$-means clustering in the bulk solution was selected using variational cross-validation with a generalized matrix Rayleigh quotient (GMRQ). When $k<90$, the testing score shows improvement, indicating model enhancement. Conversely, when $k>90$, the testing score decreases, indicating the onset of overfitting. In this study, $k=90$ was chosen as the optimal number of centers.
}
\label{GMRQ-bulk}
\end{figure}

\newpage
\begin{figure}[H]
\centering
\vspace{5mm}
\includegraphics[width=0.6 \textwidth]{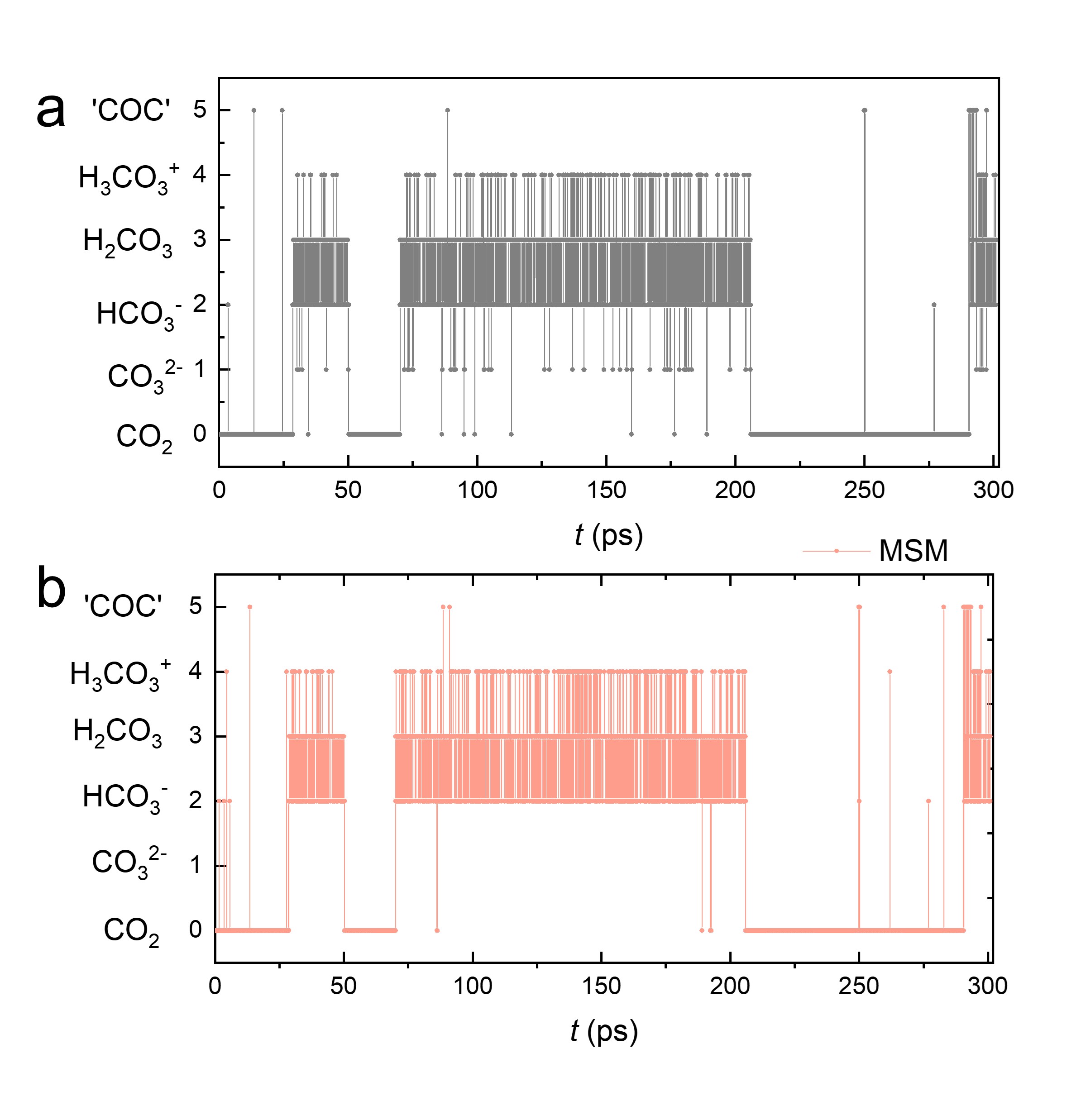}
\caption{Comparison of the carbon speciations obtained from ($\mathbf{a}$) a long AIMD trajectory  and ($\mathbf{b}$) the MSM.
Our MSM accurately identifies all the carbon species. The only slight discrepancy is that our MSM classifies CO$_3^{2-}$ and HCO$_3^-$ ions into the same group. This grouping can be attributed to the short lifetime of the CO$_3^{2-}$ ion, which leads to similarities in its dynamics with the HCO$_3^-$ ion. The long AIMD trajectory is from [13].}
\label{carbon-bulk}
\end{figure}

\newpage
\begin{figure}[H]
\centering
\vspace{5mm}
\includegraphics[width=0.6 \textwidth]{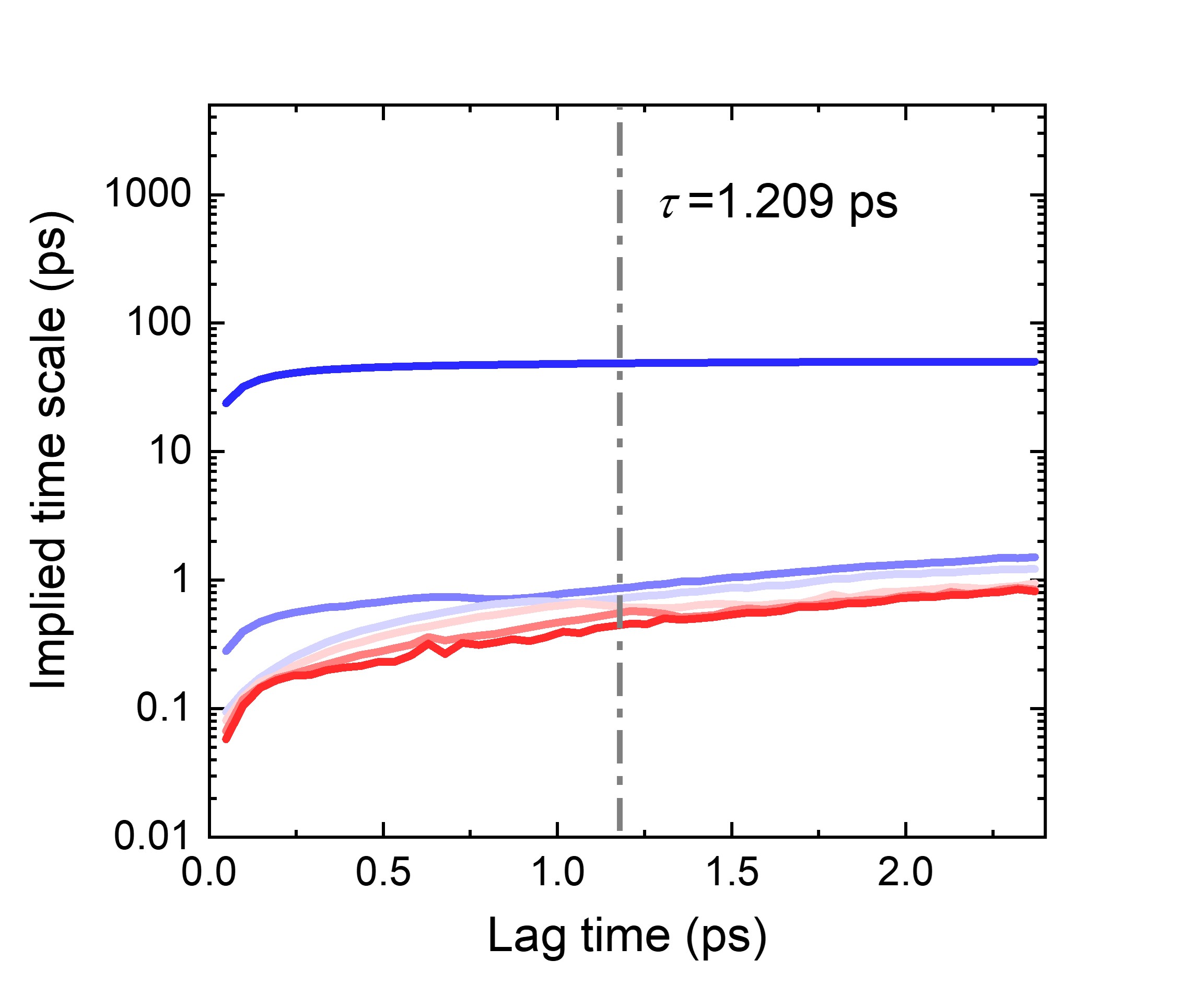}
\caption{Implied time scale analysis for the Markov state model (MSM) of the bulk solution. The implied time scales, shown by solid lines with different colors, were plotted against the lag time. We chose a lag time of $\tau=1.209$ ps where the slowest timescales level off, indicating that the MSM reaches Markovian at lag time equal to or greater than $\tau$. }
\label{ITS-bulk}
\end{figure}
\newpage
\begin{figure}[H]
\centering
\vspace{5mm}
\includegraphics[width=1.0 \textwidth]{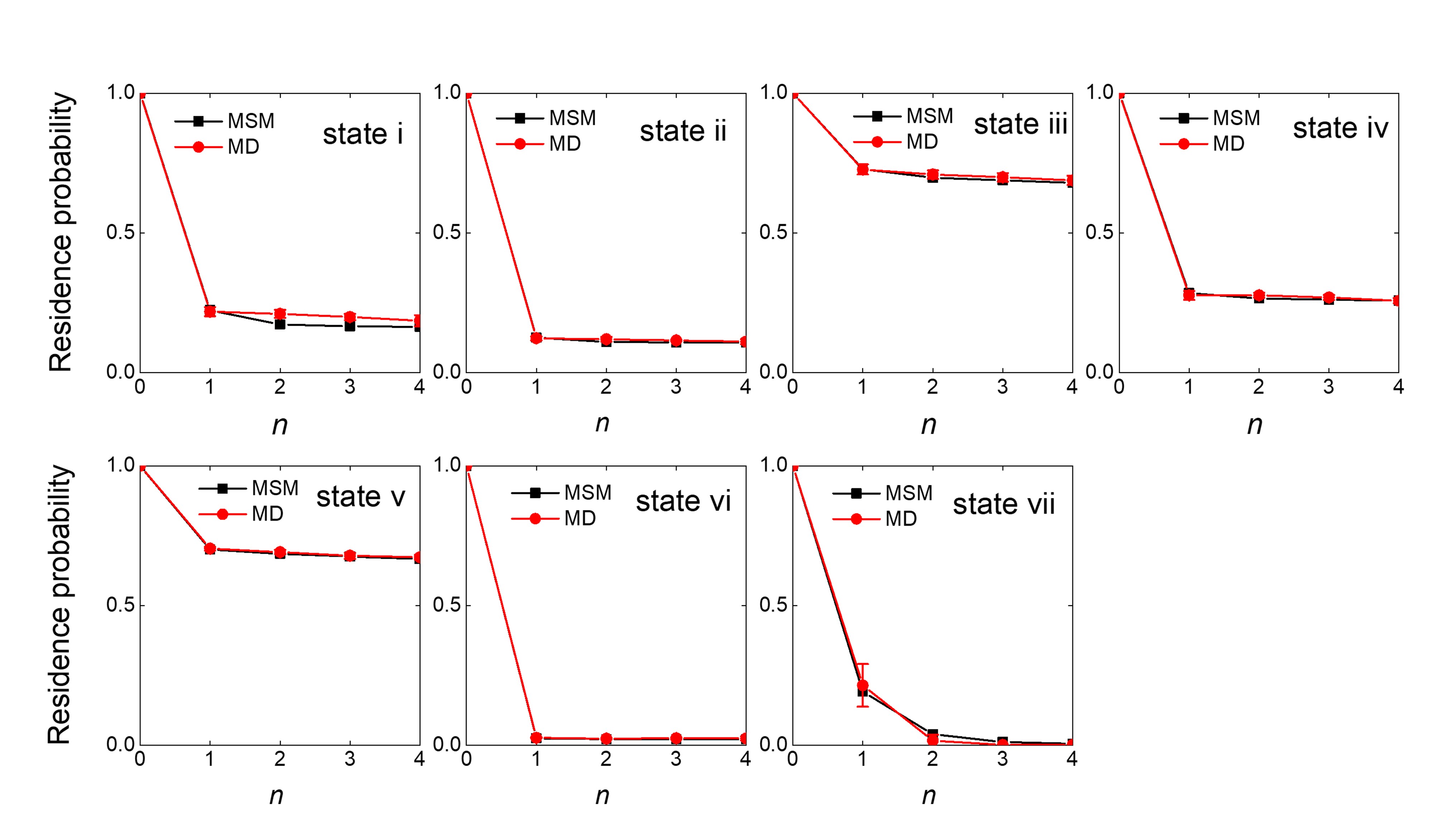}
\caption{Chapman-Kolmogorov (CK) test for the MSM of the bulk solution. 
The MSM was validated by comparing the residence probabilities obtained from the MSM and original AIMD trajectories, while the system is remaining in the same macrostates after $n$ times of the lag time ( $\tau=1.209$ ps). The residence probability from the MSM was obtained by propagating its transition probability matrix. The residence probability from AIMD was obtained by counting the original MD trajectories.
The error bars represent the standard errors of the original MD trajectories sampled by bootstrapping.
}
\label{CKtest-bulk}
\end{figure}

\newpage
\begin{figure}[H]
\centering
\vspace{5mm}
\includegraphics[width=0.5 \textwidth]{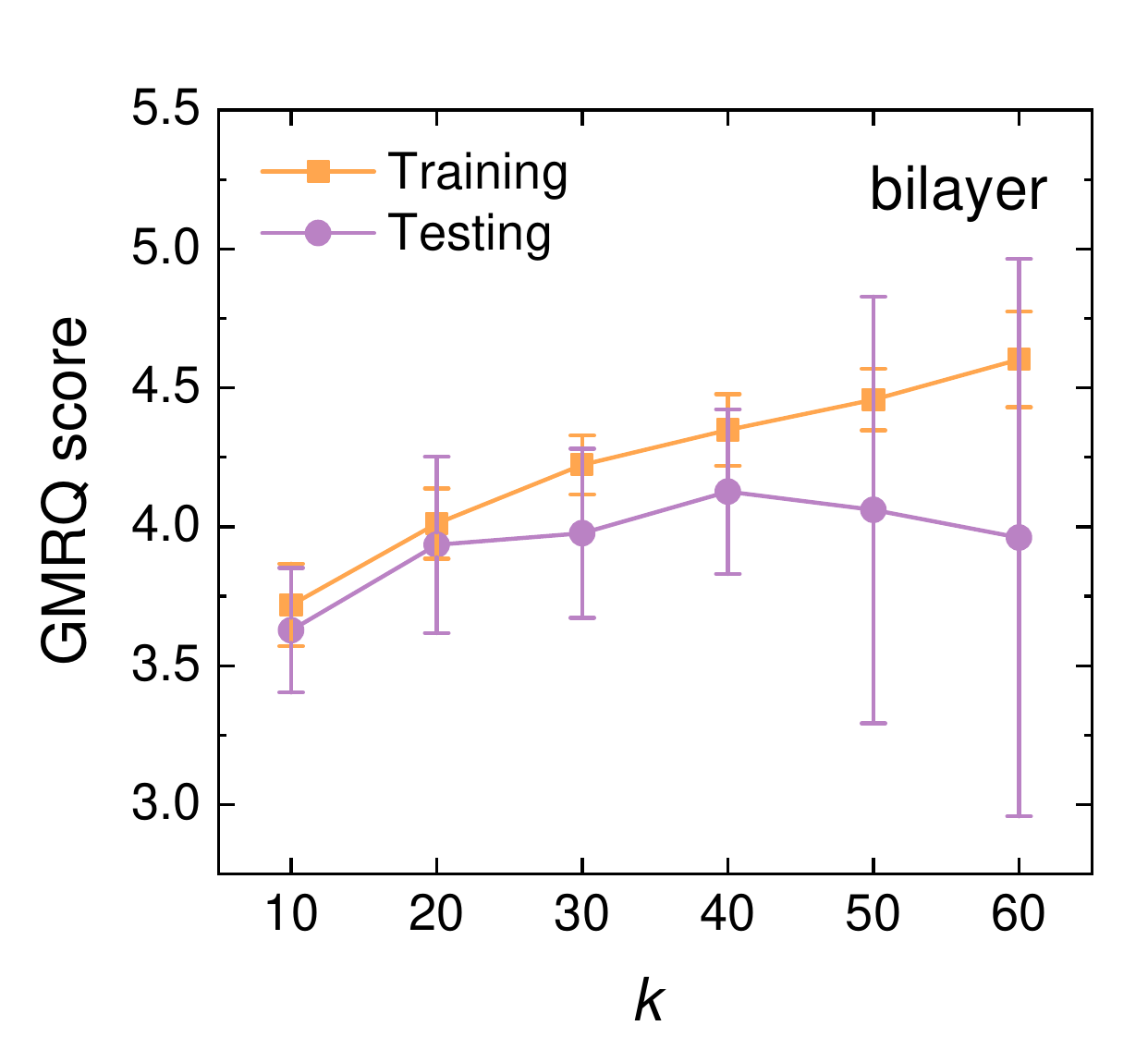}
\caption{
The number of centers for the $k$-means clustering in the bilayer solution under confinement was selected using variational cross-validation with a GMRQ.
$k=40$ was selected.  }
\label{GMRQ-bi}
\end{figure}

\newpage
\begin{figure}[H]
\centering
\vspace{5mm}
\includegraphics[width=0.5 \textwidth]{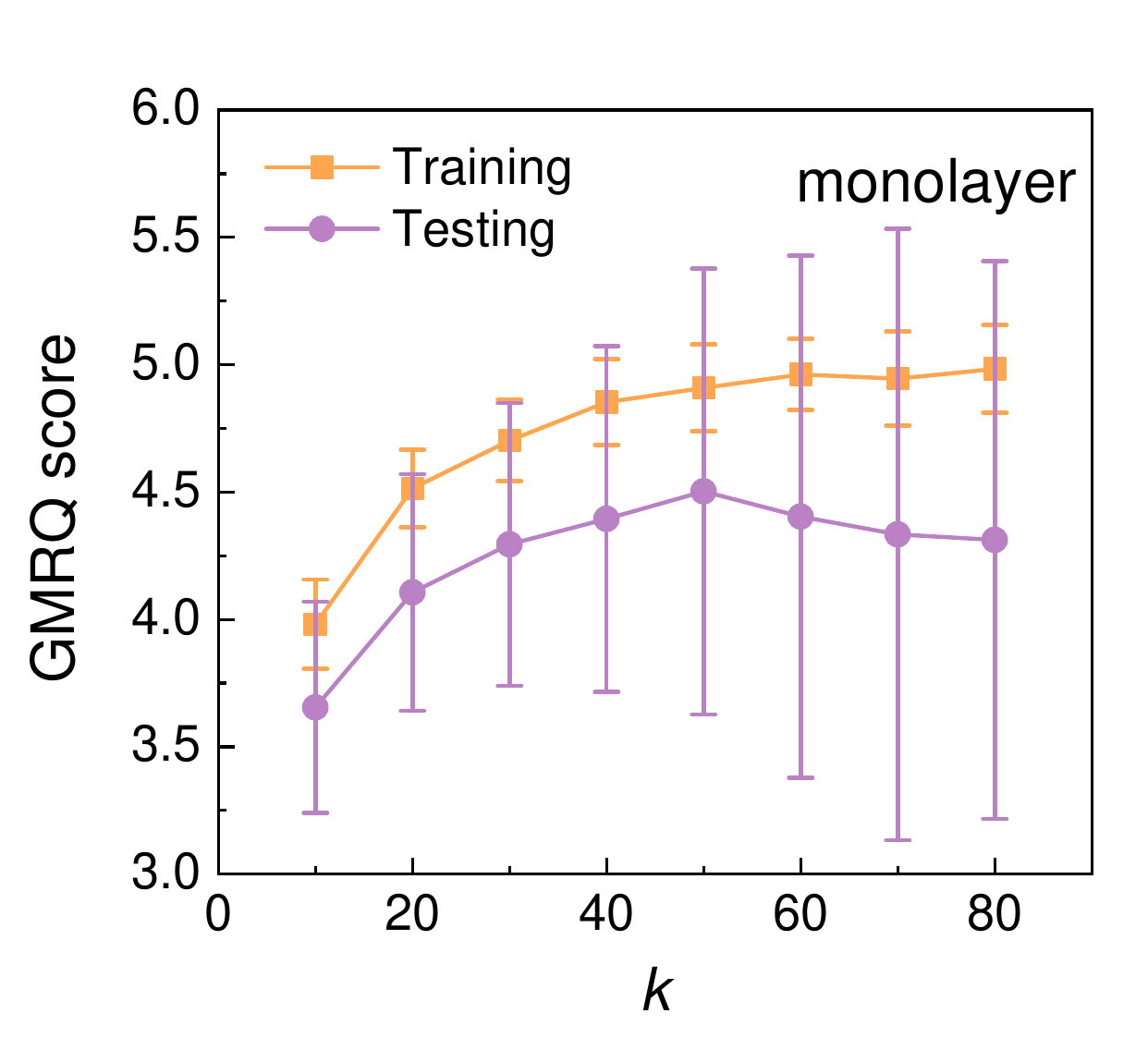}
\caption{
The number of centers for the $k$-means clustering in the monolayer solution under confinement was selected using variational cross-validation with a GMRQ.
$k=50$ was selected.  
}
\label{GMRQ-mo}
\end{figure}

\newpage
\begin{figure}[H]
\centering
\vspace{5mm}
\includegraphics[width=0.6 \textwidth]{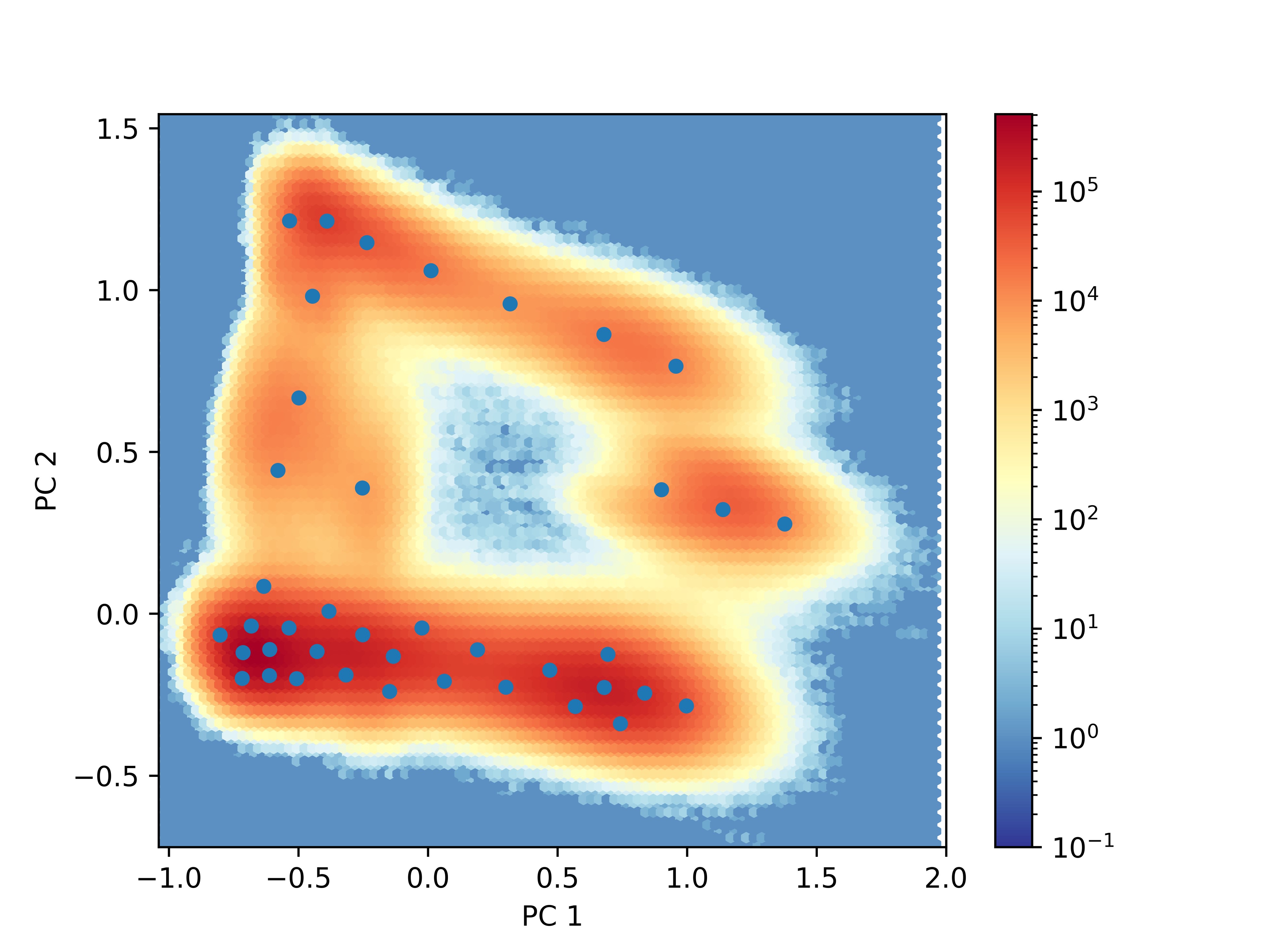}
\caption{Principal component analysis of the input $\{D_{C_iO_j}\}$ for the bilayer solution. The blue dots are the centers clustered by the $k$-means algorithm.}
\label{PCA-bilayer}
\end{figure}

\newpage
\begin{figure}[H]
\centering
\vspace{5mm}
\includegraphics[width=0.6 \textwidth]{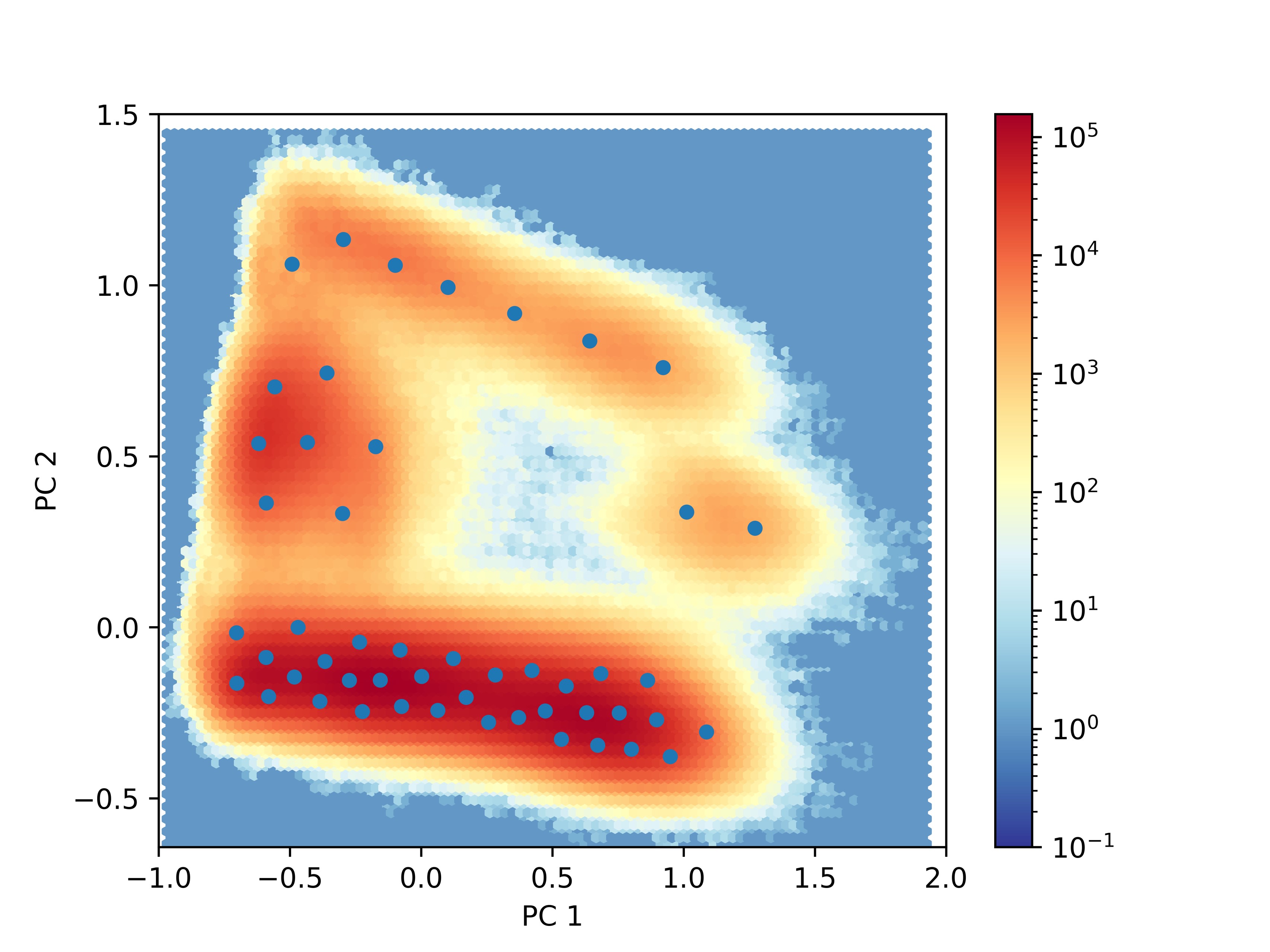}
\caption{Principal component analysis of the input $\{D_{C_iO_j}\}$ for the monolayer solution. The blue dots are the centers clustered by the $k$-means algorithm.}
\label{PCA-monolayer}
\end{figure}

\newpage
\begin{figure}[H]
\centering
\vspace{5mm}
\includegraphics[width=0.6 \textwidth]{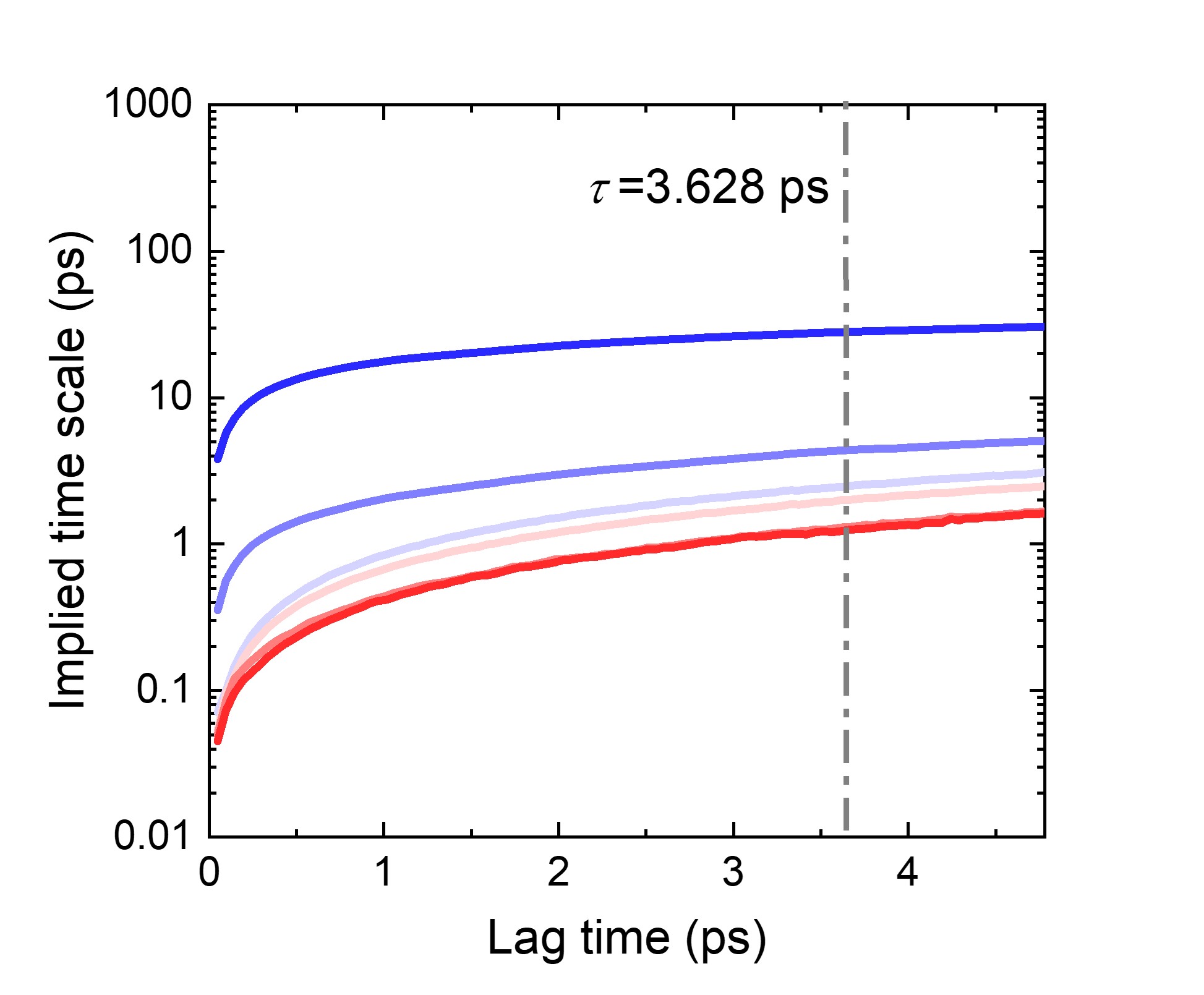}
\caption{Implied time scales analysis for the MSM of the bilayer solution. The implied time scales, shown by solid lines with different colors, were plotted against the lag time. We chose a lag time of $\tau=3.628$ ps. }
\label{ITS-bilayer}
\end{figure}

\newpage
\begin{figure}[H]
\centering
\vspace{5mm}
\includegraphics[width=0.6 \textwidth]{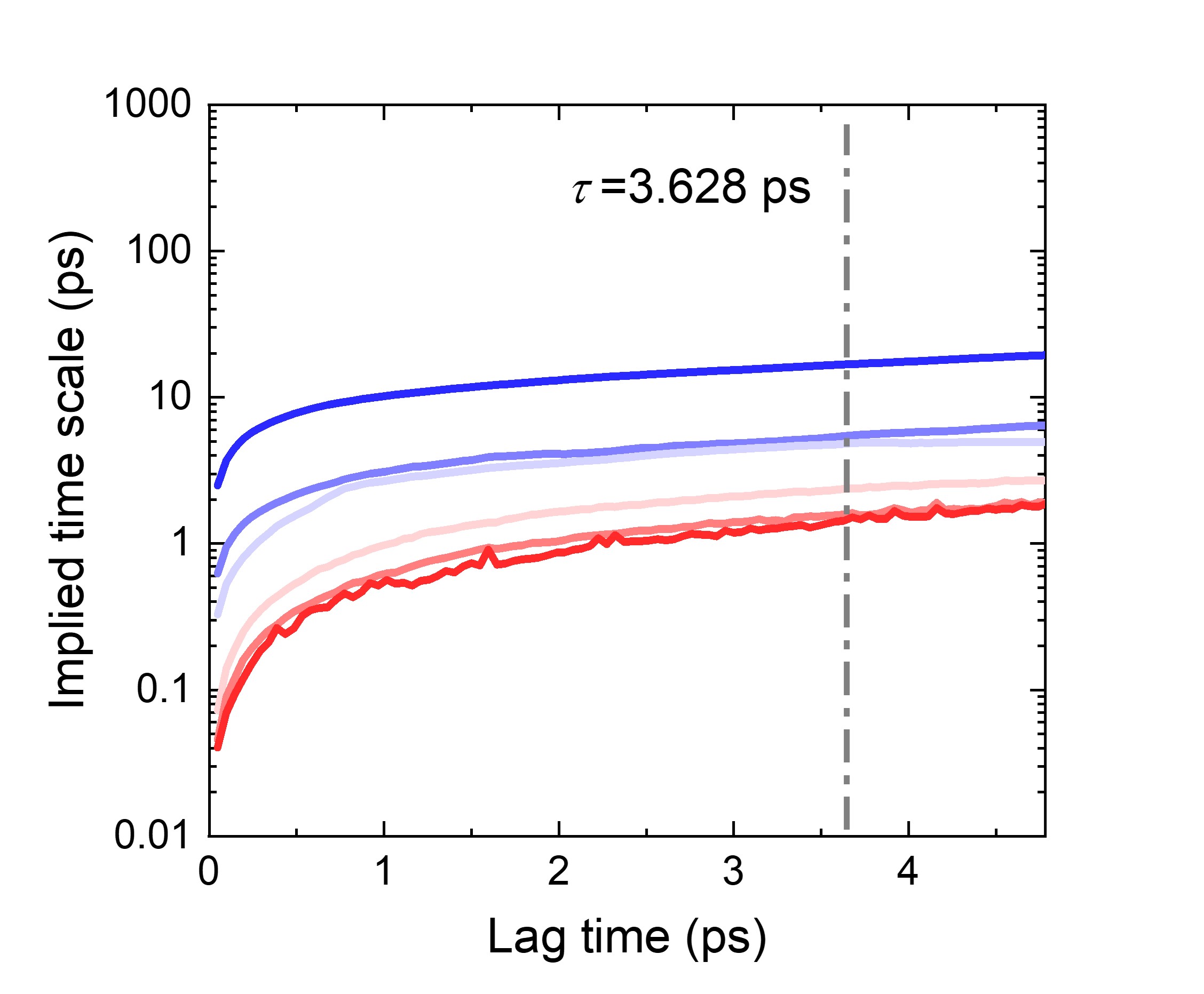}
\caption{Implied time scale analysis for the MSM of the monolayer solution. The implied time scales, shown by solid lines with different colors, were plotted against the lag time. We chose a lag time of $\tau=3.628$ ps. }
\label{ITS-monolayer}
\end{figure}

\newpage
\begin{figure}[H]
\centering
\vspace{5mm}
\includegraphics[width=1.0 \textwidth]{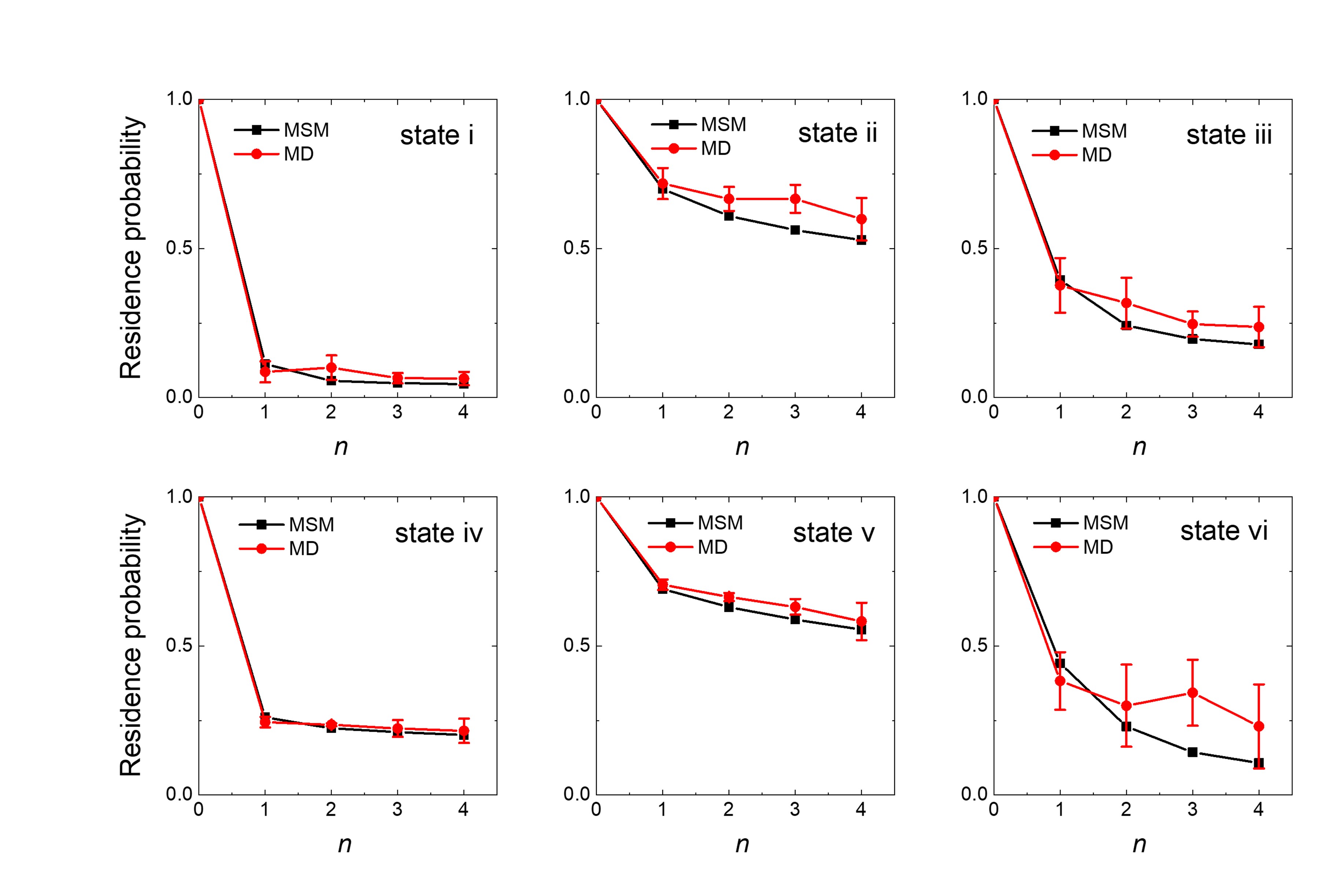}
\caption{
Chapman-Kolmogorov (CK) test for the MSM of the bilayer solution. 
The MSM was validated by comparing the residence probabilities obtained from the MSM and original AIMD trajectories, while the system is remaining in the same macrostates after $n$ times of the lag time
($\tau=3.628$ ps). The residence probability from the MSM was obtained by propagating its transition probability matrix. And the residence probability from AIMD was obtained by counting the original MD trajectories.
The error bars represent the standard errors of the original MD trajectories sampled by bootstrapping.}
\label{CKtest-bilayer}
\end{figure}

\newpage
\begin{figure}[H]
\centering
\vspace{5mm}
\includegraphics[width=1.0 \textwidth]{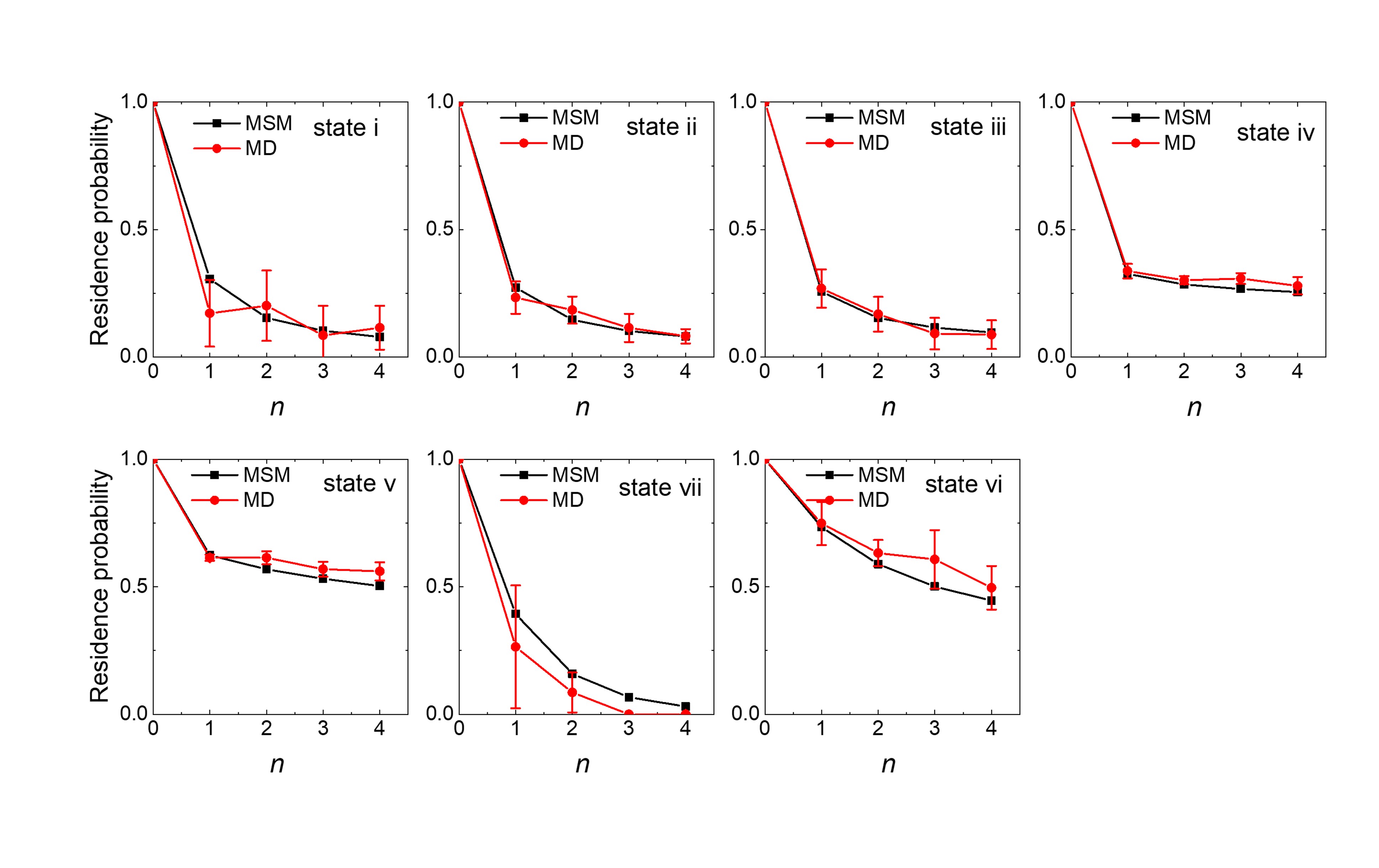}
\caption{Chapman-Kolmogorov (CK) test for the MSM of the monolayer solution. 
The MSM was validated by comparing the residence probabilities obtained from the MSM and original AIMD trajectories, while the system is remaining in the same macrostates after $n$ times of the lag time
($\tau=3.628$ ps). The residence probability from the MSM was obtained by propagating its transition probability matrix. The residence probability from AIMD was obtained by counting the original MD trajectories.
The error bars represent the standard errors of the original MD trajectories sampled by bootstrapping.}
\label{CKtest-monolayer}
\end{figure}

\newpage
\begin{figure}[H]
\centering
\vspace{5mm}
\includegraphics[width=0.6 \textwidth]{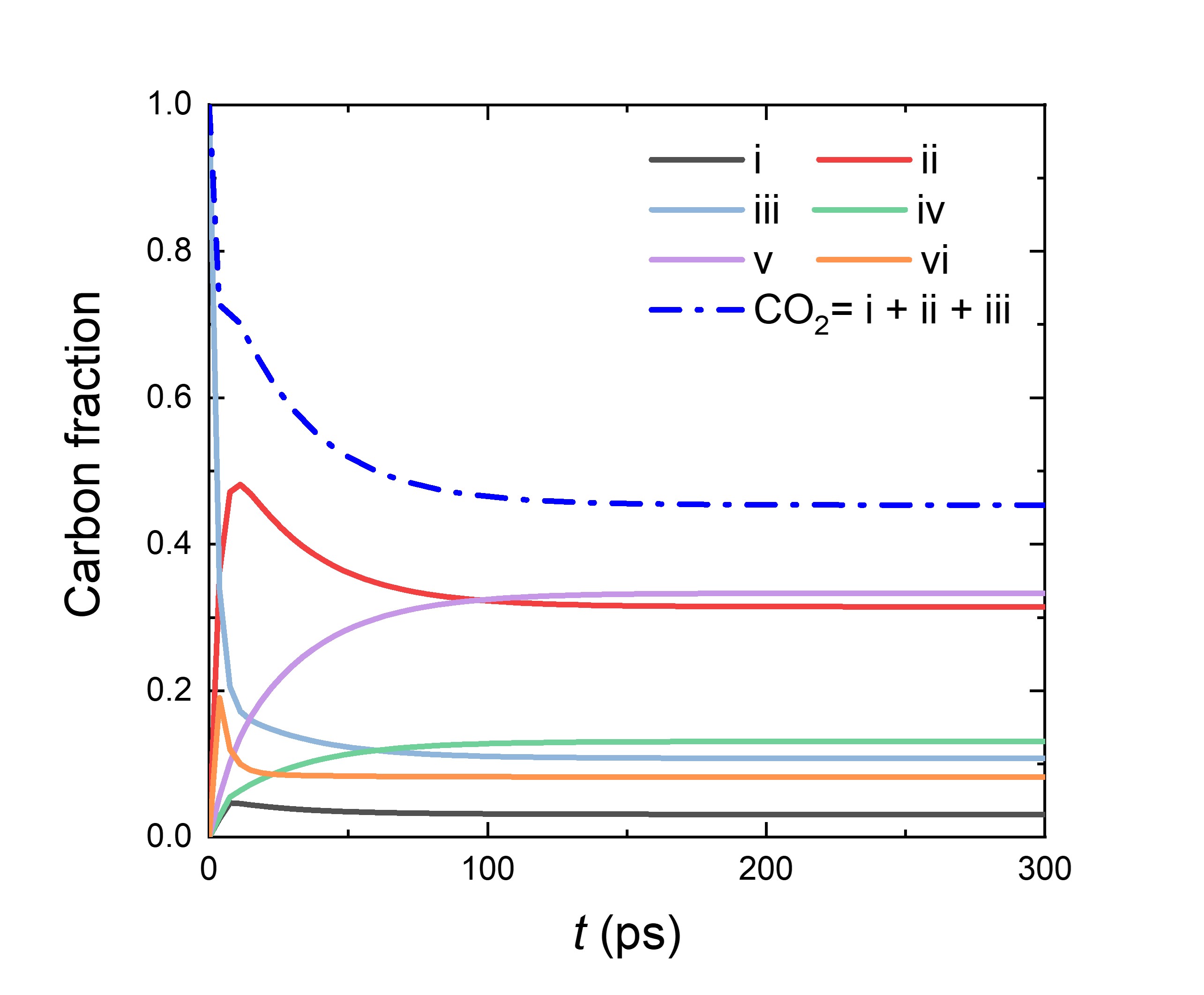}
\caption{Time evolution of carbon species in the bilayer solution obtained from the MSM.  The states are shown in Fig. 4a. The calculation started from state iii.  }
\label{evo-bilayer}
\end{figure}

\newpage
\begin{figure}[H]
\centering
\vspace{5mm}
\includegraphics[width=0.6 \textwidth]{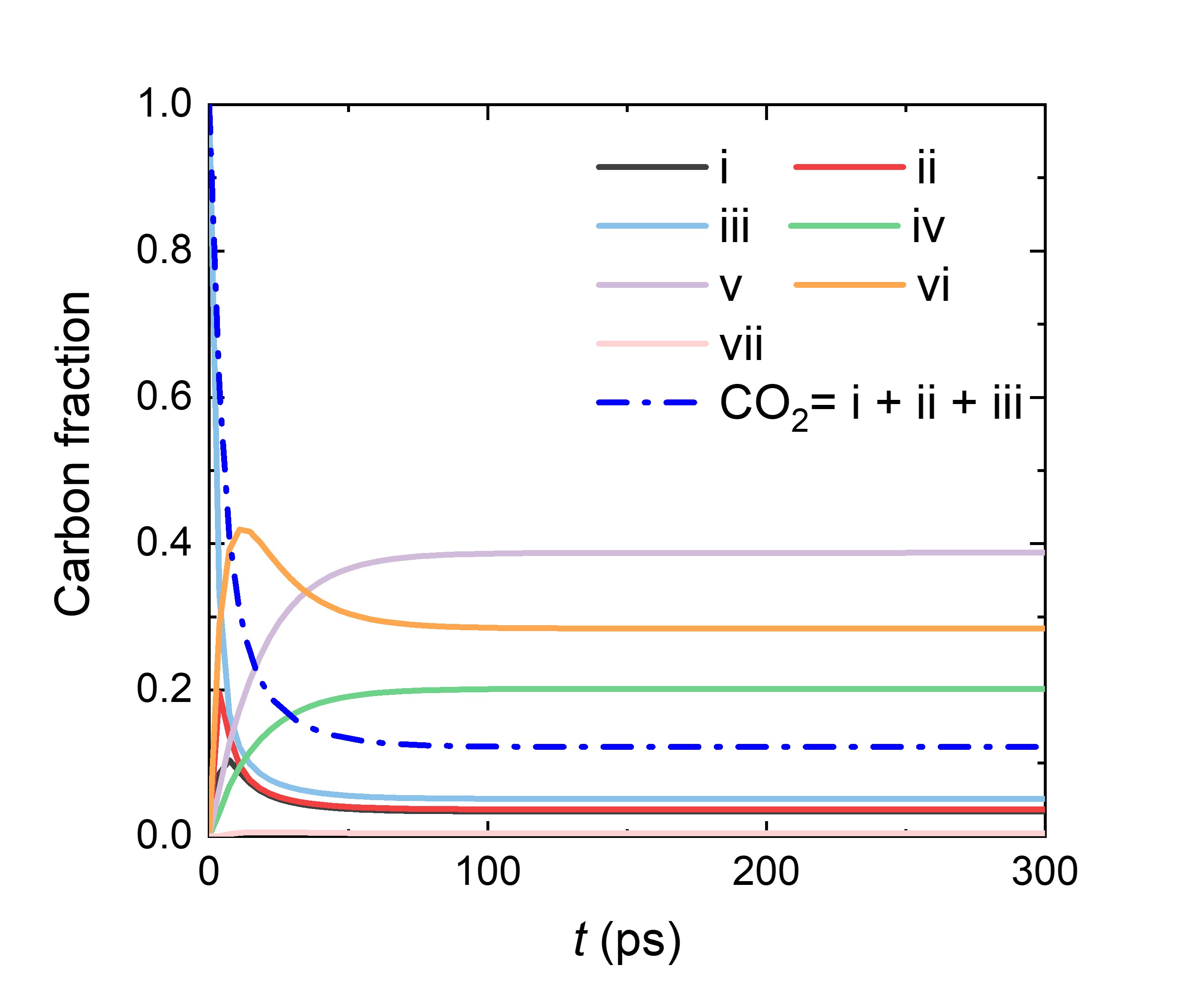}
\caption{Time evolution of carbon species in the monolayer solution obtained from the MSM.  The states are shown in Fig. 4b. The calculation started from state iii. }
\label{evo-monolayer}
\end{figure}

\newpage
\begin{figure}[H]
\centering
\vspace{5mm}
\includegraphics[width=0.9 \textwidth]{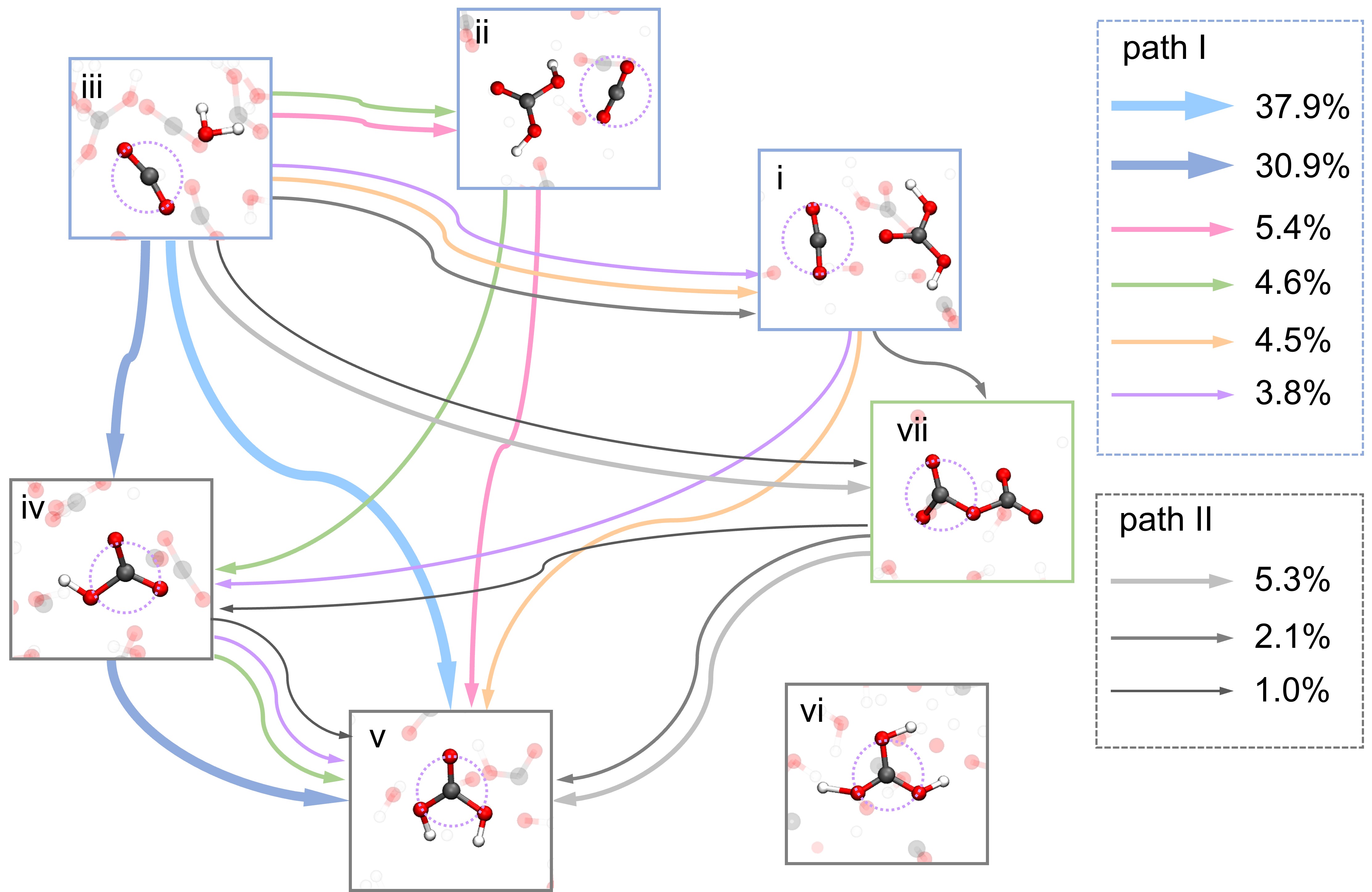}
\caption{Effective flux of carbon species in the bulk water analyzed by the transition path theory. The colored arrows represent the major paths $(> 1\% )$ from states iii to v.  
The Paths I and II are shown in Fig. 4d, and their flux percents are listed in the two dashed boxes. }
\label{flux-bulk}
\end{figure}

\newpage
\begin{figure}[H]
\centering
\vspace{5mm}
\includegraphics[width=0.9 \textwidth]{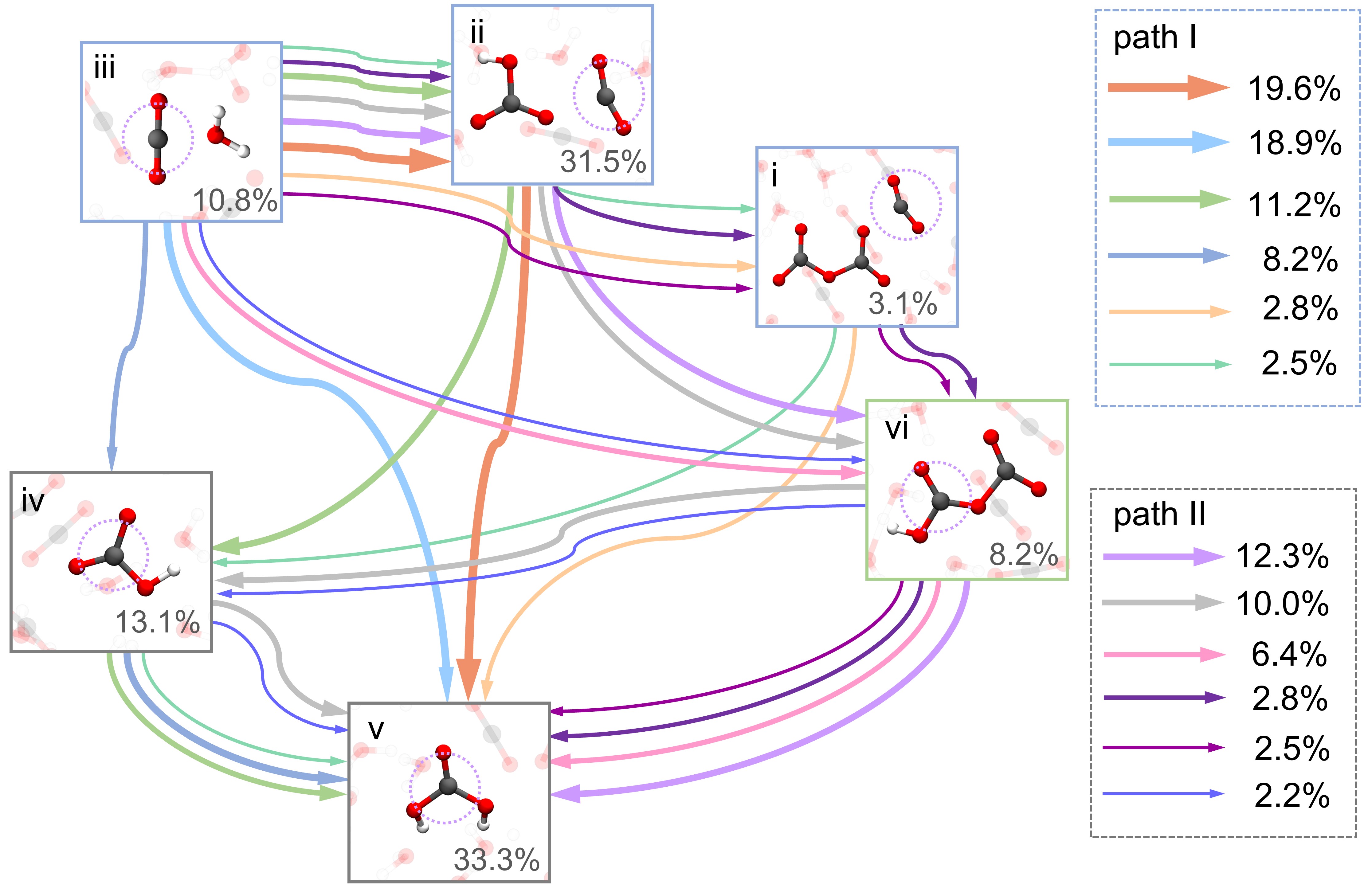}
\caption{Effective flux of carbon species in the bilayer solution analyzed by the transition path theory. The colored arrows represent the major paths $(> 1\% )$ from state iii $\longrightarrow$ v.  
The Paths I and II are shown in Fig. 4d, and their flux percents are listed in the two dashed boxes. }
\label{flux-bilayer}
\end{figure}

\newpage
\begin{figure}[H]
\centering
\vspace{5mm}
\includegraphics[width=0.9 \textwidth]{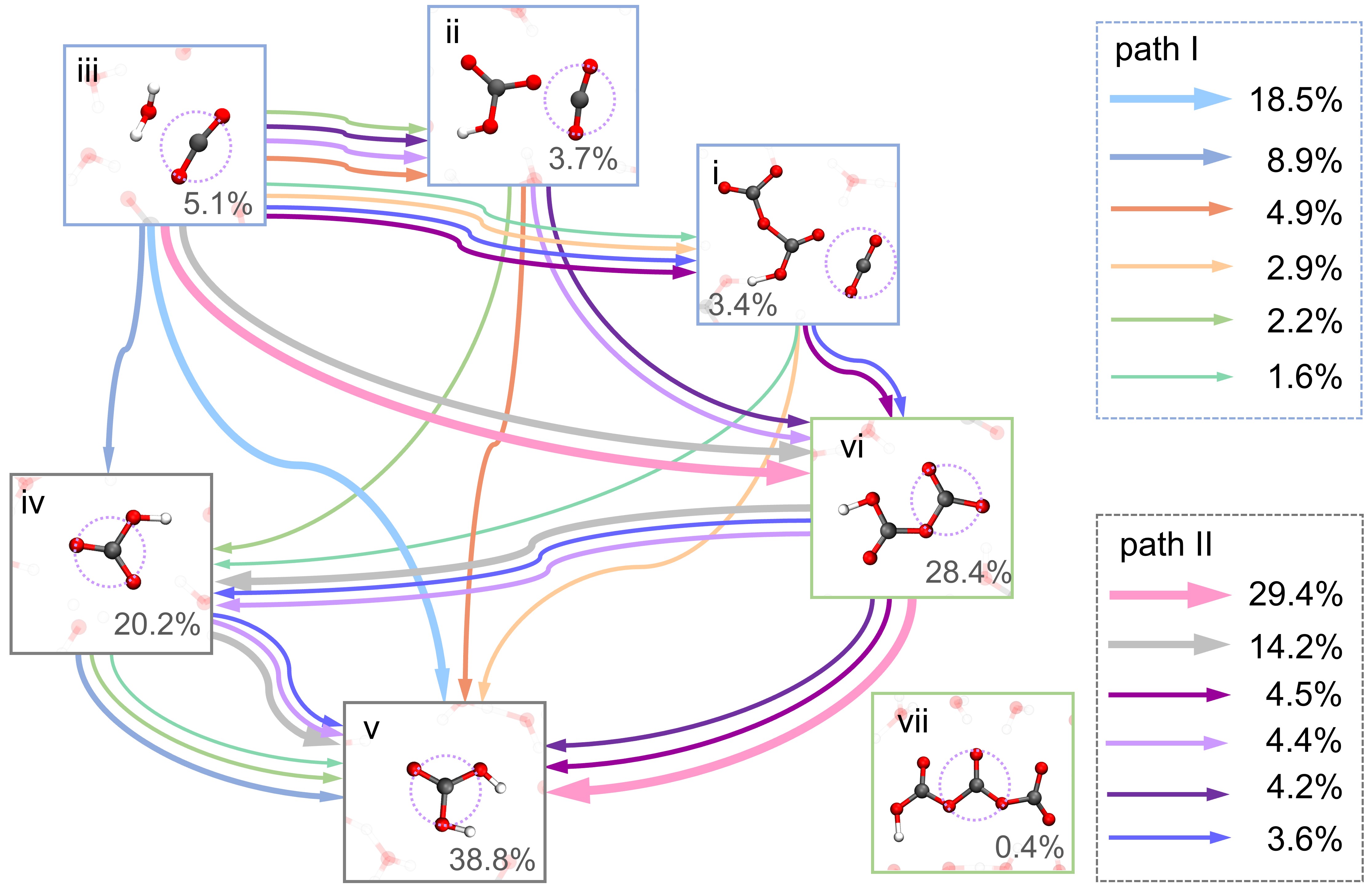}
\caption{
Effective flux of carbon species in the monolayer solution analyzed by the transition path theory. The colored arrows represent the major paths $(> 1\% )$ from states iii to v.  
The Paths I and II are shown in Fig. 4d, and their flux percents are listed in the two dashed boxes.   
}
\label{flux-monolayer}
\end{figure}

\newpage
\begin{figure}[H]
\centering
\vspace{5mm}
\includegraphics[width=1 \textwidth]{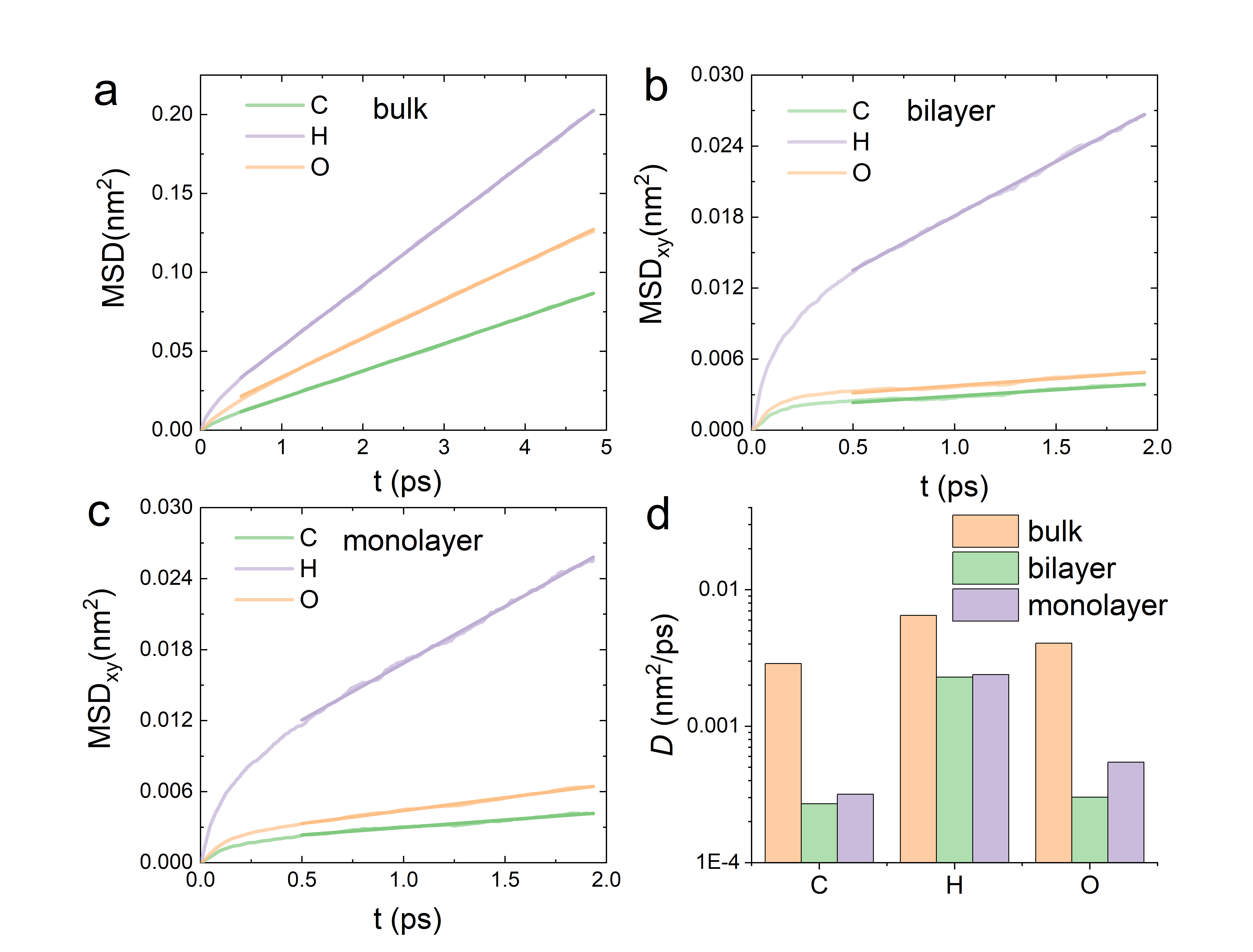}
\caption{Diffusion of hydrogen, carbon, oxygen atoms in the bulk and nanoconfined solutions. 
($\mathbf{a}$), ($\mathbf{b}$), and ($\mathbf{c}$) show the 
mean square displacements (MSD) of atoms in the bulk, bilayer, and 
monolayer solutions, respectively. ($\mathbf{d}$) Diffusion coefficients of atoms in the bulk, bilayer, and monolayer solutions. For the bulk solution, the diffusion coefficient is computed by $D=\frac{MSD}{6t}$, while under 2D confinement, the diffusion coefficient is computed by the lateral MSD as $D_{\parallel}=\frac{MSD_{xy}}{4t}$. }
\label{superionic}
\end{figure}
